\documentclass[twocolumn,showpacs,preprintnumbers,amssymb]{revtex4}

\usepackage{graphicx}
\usepackage{epsfig}
\usepackage{dcolumn}
\usepackage{bm}
\usepackage{amsmath}
\usepackage{amssymb}

\def\gsim{\lower0.5ex\hbox{$\:\buildrel >\over\sim\:$}}
\def\lsim{\lower0.5ex\hbox{$\:\buildrel <\over\sim\:$}}

\begin{document}

\preprint{UCI-TR-2007-51}

\title{Scalar FCNC and rare top decays in a two Higgs doublet model "for the top"}
\author{Itzhak Baum$^{a,b}$}
\email{setreset@tx.technion.ac.il}
\author{Gad Eilam$^b$}
\email{eilam@physics.technion.ac.il}
\author{Shaouly Bar-Shalom$^{b,c}$}
\email{shaouly@physics.technion.ac.il}
\affiliation{$^a$ Dip. di Fisica, Universit\`a di Roma “La Sapienza”, Roma, Italy\\
$^b$ Physics Department, Technion-Institute of Technology, Haifa 32000, Israel\\
$^c$ Department of Physics and Astronomy, University of California, Irvine, CA 92697, USA}

\date{\today}

\begin{abstract}
In the so called two Higgs doublet model for the top-quark (T2HDM), first suggested by Das and Kao,
the top quark receives a special status, which endows it with a naturally large mass,
and also potentially gives rise to large flavor changing neutral currents (FCNC) only in the up-quark
sector. In this paper we calculate the branching ratio (BR) for the rare decays $t\to ch$ and $h\to\bar{t}c$
($h$ is a neutral Higgs) in the T2HDM, at tree level and at 1-loop when it exceeds the tree-level.
We compare our results to predictions
from other versions of 2HDM's and find that the scalar FCNC in the T2HDM
can play a significant role in these decays.
In particular, the 1-loop mediated decays
can be significantly enhanced in the T2HDM compared to the 2HDM of types I and II,
in some instances reaching $BR \sim 10^{-4}$ which is 
within the detectable level at the LHC.
\end{abstract}


\maketitle

\section{Introduction}
The Standard Model (SM) of elementary particles has been highly successful in describing the observed and measured phenomena. It contains, however, an unexplored sector, namely, the Higgs sector. The SM also has several problems, one of which is the fermion mass hierarchy problem, especially the top quark having a much larger mass than all other quarks.

In the T2HDM which was first suggested by Das and Kao \cite{Das} 
as an extension to the SM, and which can be viewed 
as a low-energy parametrization of a more fundamental theory, the top quark receives a special status by a particular Yukawa structure which endows the top quark with a naturally large mass, while
at the same time giving rise to potentially large FCNC couplings in the up-quark sector.
Such new FCNC interactions in the up-quark sector may drive FCNC decays
such as $t\to ch$ and $h\to\bar{t}c$ that we will consider in this paper.

Previous studies \cite{bejar,t-ch LHC aguilar-saavedra} have shown that
the $BR\left(t\to ch\right)$, where $h=H^{0},h^{0},A^{0}$, could reach up
to $\sim10^{-4}$ in the 2HDM type II and in the MSSM, and about
$\sim10^{-6}$ in the 2HDM type I. In addition, the FCNC top decays
$t \to c \gamma,~cg,~cZ$ can range between $10^{-4} - 10^{-10}$ depending
on the underlying Higgs sector. In order to give a feeling of the maximal
values expected for the BR's of FCNC top rare decays to gauge-bosons and
scalars within different scalar models, we collect in table \ref{tab:rare decays} some
highlights of the results obtained in \cite{bejar} and reported in the
review \cite{t-ch LHC aguilar-saavedra}. Namely, the expected FCNC rates
within the SM and the 2HDM of types I,II and III. Note that the values are 
given with: $m_t=178$ GeV, $\overline{m_{b}}(m_{t})=2.74$ GeV and $m_{H^0}=115$ 
GeV. The branching ratio depends strongly on these parameters, especially
on $m_b$.
Note also that the
corresponding BR's for the decays into a $u$ quark instead of a $c$ quark
are a factor $\left|V_{ub}/V_{cb}\right|^2\simeq 0.0088$ smaller in the SM
and in the 2HDM of types I and II.

\begin{table}
\begin{tabular}{|c|c|c|c|}
\hline
        & SM & 2HDM-I,II & 2HDM-III \vphantom{$A^{A^A}$} \tabularnewline
\hline
$BR(t\to c\gamma)$ & $\sim 5\times 10^{-14}$ & $\lsim 10^{-9}$ & $\lsim 10^{-6}$ \vphantom{$A^{A^A}$} \tabularnewline
\hline
$BR(t\to cg)$      & $\sim 5\times 10^{-12}$ & $\lsim 10^{-8}$ & $\lsim 10^{-4}$ \vphantom{$A^{A^A}$} \tabularnewline
\hline
$BR(t\to cZ)$      & $\sim 1\times10^{-14}$ & $\lsim 10^{-10}$ & $\lsim 10^{-7}$ \vphantom{$A^{A^A}$} \tabularnewline
\hline
$BR(t\to ch)$      & $\sim 6\times 10^{-15}$ & $\lsim 10^{-5}$ & $\lsim 10^{-3}$ {\tiny (tree level)} \vphantom{$A^{A^A}$} \tabularnewline
\hline
\end{tabular}
\caption{\label{tab:rare decays}Top quark rare decays into $c\gamma$, $cg$, $cZ$, $ch$, within various models: the SM, 
the 2HDM of type I, II and type III. Our results are consistent with \cite{t-ch LHC aguilar-saavedra, t-ch SM, h-tc SM arhrib}.}
\end{table}

The estimated LHC discovery limit for $t\to ch$ is $BR(t \to ch) 
\sim 5\times10^{-5}$ \cite{t-ch LHC aguilar-saavedra}
for an integrated luminosity of $100~fb^{-1}$.
In the SM, this decay has a vanishingly small branching ratio $BR_{SM}(t \to cH^0)\sim 6\times 10^{-15}$ \cite{t-ch LHC aguilar-saavedra,t-ch SM,h-tc SM arhrib}, which is not accessible at the upcoming
Large Hadron Collider (LHC). Thus the rare FCNC decays $t\to ch$ and $h\to\bar{t}c$ are
extremely sensitive probes of new physics in the scalar sector.

In this work we will explore these rare decay channels, $t\to ch$ and $h\to\bar{t}c$, within the parameter space of the T2HDM, at the 1-loop level and, when allowed, also at the tree-level. We will focus on regions of the parameter space in which the $BR\left(t\to ch\right)$ can exceed the detection limit of the LHC, and also on regions where the $BR\left(t\to ch\right)$ and $BR\left(h\to\bar{t}c\right)$ can be enhanced significantly compared to other 2HDM.

The paper is organized as follows: in sections II and III we describe the main features of the T2HDM relevant 
for our analysis and in section IV we shortly discuss the constraints on the parameter space of the model. 
In section V we outline our analytical derivation and in section VI we give our numerical results. 
In section VII we summarize. In Appendix A we list the required Feynman rules, in Appendix B 
we give the 1-loop amplitudes, in Appendix C we define the 1-loop integrals and in Appendix D we 
derive the total Higgs width in the models considered.  

\section{\label{sec:t2hdm} The two higgs doublets model ``for the top"}

In the T2HDM the second Higgs field couples only to the top-quark, while the first Higgs field couples to all other quarks \cite{Das}:
\begin{align}
\mathcal{L}_{Y}= & -\bar{Q}_{Li}\Phi_{1}F_{ij}d_{Rj}-\bar{Q}_{Li}\tilde{\Phi}_{1}G_{ij=1,2}\left(\begin{array}{c}
u\\c\end{array}\right)_{R}\nonumber\\
& -\bar{Q}_{Li}\tilde{\Phi}_{2}G_{i3}t_{R}+h.c.\mbox{ ,}
\label{eq:LY}\end{align}
where: $i,j=1,2,3$ are flavor indices, $L(R)\equiv\left(1-(+)\gamma^{5}\right)/2$
are the chiral left (right) projection operators, $f_{L(R)}=L(R)f$
are left(right)-handed fermion fields, $Q_{L}$ is the left-handed
$SU(2)$ quark doublet and $F,G$ are general $3\times3$
Yukawa matrices in flavor space. Also, $\Phi_{1,2}$ are the Higgs doublets:
\begin{align*}
\Phi & =\left(\begin{array}{c}
\Phi^{+}\\
\frac{v+\Phi^{0}}{\sqrt{2}}\end{array}\right),\quad\tilde{\Phi}=\left(\begin{array}{c}
\frac{v^{*}+\Phi^{0*}}{\sqrt{2}}\\
-\Phi^{-}\end{array}\right).
\end{align*}

The Yukawa texture of (\ref{eq:LY}) can be realized in terms of a $Z_2$-symmetry under which the fields transform as follows \cite{Das}:
\begin{eqnarray}
 & \Phi_{1}\to-\Phi_{1}, & \quad d_{R}\to-d_{R},\quad u_{R}\to-u_{R}\;(u=u,c),\nonumber \\
 & \Phi_{2}\to+\Phi_{2}, & \quad Q_{L}\to+Q_{L},\quad t_{R}\to+t_{R}.\label{eq:z2}
 \end{eqnarray}

The Higgs potential is a general 2HDM one \cite{HHG}:
\begin{align}
\mathcal{L}_{H}= & \lambda_{1}\left(\Phi_{1}^{+}\Phi_{1}-v_{1}^{2}/2\right)^{2}+\lambda_{2}\left(\Phi_{2}^{+}\Phi_{2}-v_{2}^{2}/2\right)^{2}\nonumber\\ &+\lambda_{3}\left[\left(\Phi_{1}^{+}\Phi_{1}-v_{1}^{2}/2\right)+\left(\Phi_{2}^{+}\Phi_{2}-v_{2}^{2}/2\right)\right]^{2}\nonumber \\
&+\lambda_{4}\left[\left(\Phi_{1}^{+}\Phi_{1}\right)\left(\Phi_{2}^{+}\Phi_{2}\right)-\left(\Phi_{1}^{+}\Phi_{2}\right)\left(\Phi_{2}^{+}\Phi_{1}\right)\right]\nonumber\\ &+\lambda_{5}\left|\Phi_{1}^{+}\Phi_{2}-v_{1}v_{2}/2\right|^{2}.\label{eq:higgs potential}
\end{align}

\noindent where we have included the term $\propto \lambda_{5}$ which softly
breaks the $Z_2$-symmetry in (\ref{eq:z2}) and which can also give rise to CP-violation
\cite{georgi soft CP and Z2}.

The top-quark acquires a mass term primarily from
the second Higgs vacuum expectation value (VEV), which we will choose
to be much larger than the first Higgs VEV:
\begin{align}
\tan\beta\equiv\frac{v_{2}}{v_{1}} & \gg1 ~\label{tanbeta}.
\end{align}
\textit{Eq.~\ref{tanbeta} above is the working assumption of the T2HDM.}

The particular Yukawa structure of (\ref{eq:LY}) gives rise to various other interesting features of the T2HDM:

\begin{itemize}

\item {\bf Enhanced $H^{+}\bar{c}b$ coupling:} The $H^{+}\bar{c}b$ interaction term is
naively enhanced by a factor of $V_{tb}/V_{cb}$ compared to other 2HDM's, where $V$
is the CKM matrix. This property which
motivated the analysis in \cite{soni 15 Z-bs,soni 21 most calcs}, also
motivated the present work since, as we shall later see, the 1-loop FCNC decays $t \to c h$ and $h \to \bar{t}c$
are expected to be enhanced, due to this large $H^{+}cb$ coupling,
naively by this factor compared to e.g., the 2HDM of type II.

\item {\bf Tree-level FCNC couplings in the up-quark sector:}
While there are no tree-level FCNC interactions in the down-quark sector
(as for the example in the case of the type III 2HDM \cite{atwood reina soni bound neutral}),
there are a-priori ${\cal O}(1)$ FCNC $htc$ and $htu$ couplings.

\item {\bf Enhanced couplings to up quarks:}
The couplings of the three neutral scalars ($h \equiv H^{0},~h^{0},~or~ A^{0}$) to all
the quarks except for the top quark, increase with $\tan\beta$.
For example, the $h cc$ coupling is $\propto m_c \tan\beta$ in the T2HDM as opposed
to being $\propto m_c/\tan\beta$ in e.g., the type II 2HDM which also underlies the MSSM.
Since $\tan\beta >> 1$ is the working assumption of the T2HDM, one expects
a large enhancement of the $h cc$ coupling in the T2HDM. This
motivated the work in \cite{soni 13 3jet}.

\end{itemize}

\section{\label{sec:Yukawa}Yukawa interactions and the Theoretical setup}

A detailed derivation of all Yukawa interactions in the physical mass
basis can be found in \cite{thesis} (see also \cite{soni 34 best}).
We use the results of \cite{thesis}, which are also summarized as
a list of scalar-quark-quark Feynman rules in
Appendix \ref{app:Feynman-rules}.

Below we highlight only the new Yukawa interactions that will be the
focus of our analysis. In particular, as mentioned above, in the T2HDM the $H^+ \bar c b$
interaction is different from the typical 2HDM scenario:
\begin{widetext}
\begin{align}
\mathcal{L}_{H^+cb} & = \frac{g}{\sqrt{2}m_{W}}H^{+}\bar{c}\left[\tan\beta VM_{d}R+\left(-M_{u}\tan\beta+\Sigma^{\dagger}\left(\tan\beta+\cot\beta\right)
\right)VL\right]_{cb}b \nonumber \\
 & \sim\frac{g}{\sqrt{2}m_{W}}H^{+}\bar{c}\left[\tan\beta V_{cb}m_{b}R+m_{c}\left(-\tan\beta V_{cb}+\xi^{*}\left(\tan\beta+\cot\beta\right)V_{tb}\right)L\right]b \label{LHcb} \mbox{ ,}
 \end{align}

\noindent where ~$M_d=diag(m_d,m_s,m_b)$, $M_u=diag(m_u,m_c,m_t)$ and $\Sigma$ is a new mixing matrix in the up-quark sector that can be parametrized
as \cite{thesis}:
\begin{align}
\frac{\Sigma}{m_{t}} & =\left(\begin{array}{ccc}
\frac{m_{u}}{m_{t}}\epsilon_{ct}^{2}\left|\xi'\right|^{2}\left(1-\left|\epsilon_{ct}\xi\right|^{2}\right) & \frac{m_{u}}{m_{t}}\epsilon_{ct}^{2}\xi'^{*}\xi\sqrt{1-\left|\epsilon_{ct}\xi\right|^{2}} & \frac{m_{u}}{m_{t}}\epsilon_{ct}\xi'^{*}\left(1-\left|\epsilon_{ct}\xi\right|^{2}\right)\sqrt{1-\left|\epsilon_{ct}\xi'\right|^{2}}\\
\epsilon_{ct}^{3}\xi^{*}\xi'\sqrt{1-\left|\epsilon_{ct}\xi\right|^{2}} & \epsilon_{ct}^{3}\left|\xi\right|^{2} & \epsilon_{ct}^{2}\xi^{*}\sqrt{1-\left|\epsilon_{ct}\xi\right|^{2}}\sqrt{1-\left|\epsilon_{ct}\xi'\right|^{2}}\\
\epsilon_{ct}\xi'\left(1-\left|\epsilon_{ct}\xi\right|^{2}\right)\sqrt{1-\left|\epsilon_{ct}\xi'\right|^{2}} & \;\epsilon_{ct}\xi\sqrt{1-\left|\epsilon_{ct}\xi\right|^{2}}\sqrt{1-\left|\epsilon_{ct}\xi'\right|^{2}} & \left(1-\left|\epsilon_{ct}\xi\right|^{2}\right)\left(1-\left|\epsilon_{ct}\xi'\right|^{2}\right)\end{array}\right),\label{eq:sigma}
\end{align}
\end{widetext}

\noindent where $\epsilon_{ct} \equiv m_c/m_t$ and $\xi,\xi^\prime$ are dimensionless parameters
naturally of ${\cal O}(1)$.

As we can see from (\ref{LHcb}), the $H^{+}\bar{c}b$ vertex has terms proportional
to $V_{cb}$ which are common to other 2HDM's, but has
an additional term which is not CKM suppressed and is proportional to
$\left(\tan\beta+\cot\beta\right)\times\left(\Sigma^{\dagger}V\right)_{cb}\sim m_{c}\xi^{*} \tan\beta V_{tb}$.
As we shall see below, this apparent enhancement to $H^{+}\bar{c}b$ coupling will drive the main contribution to the 1-loop diagrams with internal $H^{+}$ and $b$.

The $H^{+}\bar{t}b$ vertex also receives an additional term within the T2HDM:
\begin{align}
\mathcal{L}_{H^+tb}\sim
 &\frac{g}{\sqrt{2}m_{W}}H^{+}\bar{t}\left\{ \tan\beta V_{tb}m_{b}R+\left[m_{t}V_{tb}\cot\beta  \vphantom{\left|\xi\right|^2}\right.\right.\nonumber\\
 &\left.\left.-m_{t}V_{tb}\epsilon_{ct}^{2}\left(\left|\xi\right|^{2}+\left|\xi'\right|^{2}\right)\left(\tan\beta+\cot\beta\right)\right]L\right\}b.\label{eq:H+tb}
\end{align}

We see, however, that this new contribution to the
$H^+ tb$ coupling is sub-leading and vanishes in the limit $\xi,\xi'\to0$,
in which case the $H^{+}\bar{t}b$ interaction in \eqref{eq:H+tb} converges to
that of a 2HDM types I and II.

As for the neutral Higgs sector, there is no \textit{a priori} distinction between $h^{0}$ and $H^{0}$ other than the rotation angle $\alpha$.
In particular, the $h^0 tc$ and $H^0 tc$ Yukawa interactions are:
\begin{align}
\mathcal{L}_{h^0tc} & \sim h^{0}\bar{t}\left[-\frac{g}{2m_{W}} \left(\frac{\cos\alpha}{\sin\beta}+\frac{\sin\alpha}{\cos\beta}\right) m_{c}\xi \left( R+\epsilon_{ct} L\right)\right]c~,
\end{align}
\begin{align}
\mathcal{L}_{H^0 tc} & \sim H^{0}\bar{t}\left[\frac{g}{2m_{W}} \left(-\frac{\sin\alpha}{\sin\beta}+\frac{\cos\alpha}{\cos\beta}\right) m_{c}\xi \left( R+\epsilon_{ct} L\right)\right]c \label{H0tc}~,
\end{align}

\noindent where we have used the off-diagonal terms of $\Sigma$ from
Eq.~\ref{eq:sigma}, neglecting terms of order $\epsilon_{ct}^{2}$ (recall $\epsilon_{ct}=m_{c}/m_{t}$) for which
$\Sigma_{tc}\sim m_{c}\xi$ and $\left(\Sigma^{\dagger}\right)_{tc}\sim m_{c}\epsilon_{ct}\xi$.

For arbitrary $\alpha$ and $\beta$, the above FCNC interactions will lead to both
$t\to cH^{0}$ and $t\to ch^{0}$ (or $H^0 \to \bar{t}c$ and $h^0 \to \bar{t}c$ if $m_t < m_{H^0},m_{h^0}$) decays at tree level.
One can, however, eliminate one of the two tree-level $h^0tc$ or $H^0tc$ couplings
by choosing a specific direction with respect to the mixing angles $\alpha$ and $\beta$ of the neutral Higgs sector.
For example,
the tree-level $H^{0}\bar{t}c$ coupling can be eliminated if one of the
following two conditions is satisfied:

\begin{enumerate}

\item $\xi=0$.
 \item $-\frac{\sin\alpha}{\sin\beta}+\frac{\cos\alpha}{\cos\beta}=0$ implying $\alpha=\beta+n\pi$.
\end{enumerate}

Without loss of generality, let us consider the case in which either the $h^0tc$ or the $H^0tc$ coupling vanishes at tree-level.
For definiteness, we will adopt the second choice above, $\alpha=\beta$, which sets:
\begin{align}
\mathcal{L}^{\alpha=\beta}_{H^0tc} = 0~, \end{align}

\noindent and gives:

\begin{align}
\mathcal{L}^{\alpha=\beta}_{h^0tc} &\sim h^{0}\bar{t}\left[-\frac{g m_{c}\xi}{2m_{W}}\left(\tan\beta+\cot\beta\right)\left( R+\epsilon_{ct} L\right)\right]c \label{h0tcTL}~. \end{align}

There are several reasons which motivate an analysis of the case $\alpha=\beta$
rather than the case of generic mixing angles and for preferring this choice
over the choice $\xi=0$ which also eliminates the tree-level $H^0 tc$ coupling:

\begin{enumerate}
\item $\xi=0$ is disfavored by the analysis in \cite{soni 34 best},
as we will recapitulate in Sec. \ref{sec:par space T2HDM}.
\item $\xi=0$ diminishes the potentially enhanced term in the $H^{+}\bar{c}b$
coupling, as is evident from Eq.~\ref{LHcb}.
\item The limit $\alpha=\beta$ is a natural result of the MSSM, when the
mass of the CP-odd neutral Higgs, $A^{0}$, is large (see e.g. \cite{HHG}
in the limit $m_{A^{0}}\to\infty$). The choice $\alpha=\beta$
is widely used in the literature, partly for this reason. Thus, even though
the T2HDM setup is not natural within the MSSM, this will help us
compare our results with other existing results in different types
of 2HDM's.
\item The limit $\alpha=\beta$ sets the scalar $H^{0}$ to be SM-like,
in which case the direct bounds on the SM Higgs mass roughly apply
to $H^{0}$. Also, $H^0$ will have SM-like Yukawa couplings to quarks:
\begin{align}
\mathcal{L}_{Y_{H^0}}^{\alpha=\beta} & \supset H^{0}\bar{d}\left[-\frac{gM_{d}}{2m_{W}}\right]d+H^{0}\bar{u}
\left[-\frac{gM_{u}}{2m_{W}}\right]u.
\end{align}

\end{enumerate}

Finally, the $h^{0}\bar{t}t$ interaction, with $\alpha=\beta$, reads:
\begin{align}
\mathcal{L} \sim & h^{0}\bar{t}\left\{ \frac{g m_{t}}{2m_{W}} \left[-\cot\beta+\epsilon^{2}_{ct}\left(\left|\xi\right|^{2}+\left|\xi'\right|^{2}\right)\left(\tan\beta+\cot\beta\right)\right]R\right.\nonumber\\
& \left.+\left(h.c.\right)L\vphantom{\frac{a}{b}}\right\} t~,
\end{align}
where $\Sigma_{tt}=\left(\Sigma^{\dagger}\right)_{tt}\sim m_{t}-m_{c}\epsilon_{ct}\left(\left|\xi\right|^{2}+\left|\xi'\right|^{2}\right)$.
As in the case of $H^{+}\bar{t}b$, the term $\propto m_{t}\tan\beta$
cancels, and we are left with the usual $m_t /\tan\beta$ term plus terms that are suppressed either by $\epsilon_{ct}^{2}$
or $\cot^{2}\beta$.

\section{\label{sec:par space T2HDM}The parameter space of the
T2HDM}

The parameter space of the T2HDM was recently analyzed in \cite{soni 34 best}.
Here we recapitulate the bounds on the parameter space of T2HDM
that were derived in \cite{soni 34 best} by performing
a best fit to several experimentally measured observables mainly associated with
B-decays. The processes
that were selected were the ones that are potentially most sensitive to the
charged sector of the T2HDM. The analysis in \cite{soni 34 best} is directly relevant to the present work, and so
we list below the final results of \cite{soni 34 best} (recall that
$\xi=\left|\xi\right|e^{i\varphi_{\xi}}$):
\begin{eqnarray}
m_{H^{\pm}} & = & \left(660_{-280}^{+390}\right)\mbox{ GeV},\nonumber \\
\tan\beta & = & 28_{-8}^{+44},\nonumber \\
0.5< & \left|\xi\right| & <1,\nonumber \\
\varphi_{\xi} & = & \left(110_{-65}^{+30}\right)^{\circ},\nonumber \\
\left|\xi'\right| & \sim & 0.21,\nonumber \\
\varphi_{\xi'} & \sim & 250^{\circ}.\label{eq:bounds}
\end{eqnarray}

\noindent Note that:

\begin{itemize}
\item These values for $\tan\beta$ and $m_{H^+}$ are allowed also within
the framework of a type II 2HDM \cite{arhrib}.
\item The authors of \cite{soni 34 best} didn't consider possible constraints
on the FCNC $H^0$ - up-quark couplings of the T2HDM, coming from
the recently measured $D^0$ oscillations. We note, however, that
such contributions to the $D^0 - \bar D^0$ mass difference is suppressed by
a factor of $\left(\frac{1}{\tan\beta}\frac{m_{c}}{m_{t}}\frac{m_{H^{+}}}{m_{h^{0}}}\right)^{2}$
compared to the charged Higgs contribution that was considered in \cite{soni 34 best}, and therefore does not
impose further constraints on the FCNC parameter space of the neutral
sector other than those found in \cite{atwood reina soni bound neutral,soni 21 most calcs}.
\end{itemize}

There are also direct constraints on the neutral Higgs masses from
high-energy collider experiments \cite{PDG}:
\begin{itemize}
\item The direct bound on the SM Higgs (which also applies to $H^0$ of the T2HDM when $\alpha=\beta$) is: $m_{H^{0}}>114$ GeV.
\item The bounds on the mass of lightest neutral scalar and the charged scalar in supersymmetry
are: $m_{h}>90$ GeV and $m_{H^{+}}>80$ GeV.
\end{itemize}

\section{\label{sec:Calculations}Analytical Results}

\subsection{\label{sub:calcs Tree}The tree-level $t\to ch^{0}$ or $h^{0}\to\bar{t}c$}

As stated above, when $\alpha=\beta$ the decays $t\to ch^{0}$ or $h^{0}\to\bar{t}c$ can
proceed at tree-level (when kinematically allowed) while the corresponding decays involving $H^0$ are mediated at
1-loop.
Using the tree-level coupling $h^{0}\bar{t}c$ in Eq.~\ref{h0tcTL},
the tree-level amplitude for the process $t\to ch^{0}$ is:
\begin{align}
\mathcal{M}\left(t\to ch^{0}\right)= &\bar{u}_{c}\left[\frac{g}{2m_{W}}\left(\left(M_{u}\right)_{ct}\frac{\sin\alpha}{\cos\beta} \right.\right.\\ &\left.\left.-\Sigma_{ct}\left(\frac{\cos\alpha}{\sin\beta}+\frac{\sin\alpha}{\cos\beta}\right)\right)R+\left(h.c.\right)L\right]u_{t},\nonumber
\end{align}
where from \eqref{eq:sigma} we have:
\begin{align}
\Sigma_{ct} & =m_{t}\epsilon_{ct}^{2}\xi^{*}\sqrt{1-\left|\epsilon_{ct}\xi\right|^{2}}\sqrt{1-\left|\epsilon_{ct}\xi'\right|^{2}}\sim m_{c}\epsilon_{ct}\xi^{*},\nonumber \\
\left(\Sigma^{\dagger}\right)_{ct} & =m_{t}\epsilon_{ct}\xi^{*}\sqrt{1-\left|\epsilon_{ct}\xi\right|^{2}}\sqrt{1-\left|\epsilon_{ct}\xi'\right|^{2}}\sim m_{c}\xi^{*}.
\end{align}
Taking $\alpha=\beta$ we then obtain:
\begin{align}
\mathcal{M}\left(t\to ch^{0}\right) & =\bar{u}_{c}\frac{-g}{2m_{W}}\left(\cot\beta+\tan\beta\right)m_{c}\xi^{*}\left[\epsilon_{ct}R+L\right] u_{t} \nonumber\\ &\equiv\bar{u}_{c}\left[M_{R}R+M_{L}L\right]u_{t}.
\end{align}

The squared amplitude, summed over initial and final state polarizations is then:
\begin{eqnarray}
\underset{pol}{\sum}\left|\mathcal{M}\right|^{2}&= &2m_{c}m_{t}\left(M_{L}M_{R}^{*}+M_{R}M_{L}^{*}\right)\nonumber\\  & & +\left(m_{t}^{2}+m_{c}^{2}-m_{h}^{2}\right)\left(M_{L}M_{L}^{*}+M_{R}M_{R}^{*}\right)\nonumber \\
 &= & \frac{g^{2}m_{c}^{2}}{4m_{W}^{2}}\left(\cot\beta+\tan\beta\right)^{2}\left|\xi\right|^{2}\nonumber\\
 & & \times \left[2m_{c}m_{t}\cdot2\epsilon_{ct}+\left(m_{t}^{2}+m_{c}^{2}-
 m_{h^{0}}^{2}\right)\left(1+\epsilon_{ct}^{2}\right)\right]\nonumber \\
 &\sim & \frac{g^{2}m_{c}^{2}}{4m_{W}^{2}}\tan^{2}\beta\left|\xi\right|^{2}\left[m_{t}^{2}-m_{h^{0}}^{2}\right] \label{amp2}~.
\end{eqnarray}
where we have neglected terms of ${\cal O}(m_c^2/m_t^2)$ and of ${\cal O}(\cot\beta)$ in accordance with the working assumption
of the T2HDM, i.e., that $\tan\beta >>1$.
The width of $t\to ch^{0}$ then reads:
\begin{align}
\Gamma\left(t\to ch^{0}\right) & =4\pi\cdot\lambda^{\frac{1}{2}}\left(1,\frac{m_{c}^{2}}{m_{t}^{2}},
\frac{m_{h^{0}}^{2}}{m_{t}^{2}}\right)\cdot\frac{\underset{pol}{\sum}
\overline{\left|\mathcal{M}\right|^{2}}}{64\pi^{2}m_{t}}\nonumber\\
& \sim\frac{g^{2}\left|\xi\right|^{2}m_{t}m_{c}^{2}}{128\pi m_{W}^{2}}\tan^{2}\beta\left(1-\frac{m_{h^{0}}^{2}}{m_{t}^{2}}\right)^2,
\end{align}
where $\lambda\left(x,y,z\right)=x^{2}+y^{2}+z^{2}-2xy-2xz-2yz$ and
$\underset{pol}{\sum}\overline{\left|\mathcal{M}\right|^{2}}=\frac{1}{2}
\underset{pol}{\sum}\left|\mathcal{M}\right|^{2}$ is the squared amplitude summed over final polarizations
and averaged over the initial top polarizations.

We then obtain the following $BR\left(t\to ch^{0}\right)$ (for large $\tan\beta$):
\begin{align}
BR\left(t\to ch^{0}\right) \sim & \frac{\left|\xi\right|^{2}m_{c}^{2}}{2m_{W}^{2}}\tan^{2}\beta \left(1-\frac{m_{h^{0}}^{2}}{m_{t}^{2}}\right)^2 \left(1-\frac{m_{W}^{2}}{m_{t}^{2}}\right)^{-1} \nonumber\\
& \times \left(1-2\frac{m_{W}^{2}}{m_{t}^{2}}+\frac{m_{t}^{2}}{m_{W}^{2}}\right)^{-1},
\end{align}
where for the total top-quark decay width ($\Gamma_t$) we took (at tree-level and neglecting terms of order $m_{b}^{2}/m_{t}^{2}$ \cite{PDG}):
\begin{align}
\Gamma_t & = \Gamma\left(t\to bW^{+}\right) \nonumber\\
& \sim\frac{g^{2}m_{t}}{64\pi}\left(1-\frac{m_{W}^{2}}{m_{t}^{2}}\right)
\left(1-2\frac{m_{W}^{2}}{m_{t}^{2}}+\frac{m_{t}^{2}}{m_{W}^{2}}\right) \label{tbW}~.
\end{align}

For instance, taking $\tan\beta=28$, $\left|\xi\right|=0.8$ (compatible with the bounds in
\eqref{eq:bounds}) and $m_{h^{0}}=91$ GeV, we get (when $\Gamma_t=\Gamma(t\to bW^{+})$):
\begin{eqnarray}
BR\left(t\to ch^{0}\right) \sim 0.0077~.
\end{eqnarray}

\noindent Note that if $m_{H^{+}}< m_t$, then the top can also have an appreciable BR to
$t\to H^{+}b$ which must be taken into account in $\Gamma_t$.

The decay width for the reverse $h^{0}\to\bar{t}c$ process (corresponding to the case $m_{h^0} > m_t$)
can be obtained by applying a crossing symmetry to the
squared amplitude of the decay $t \to c h^0$ in Eq.~\ref{amp2} (see e.g., \cite{peskin}):
\begin{align}
\underset{pol}{\sum}\left|\mathcal{M}\right|^{2} (h^0 & \to \bar{t}c) =\underset{pol}{\sum}\left|\mathcal{M}\right|^{2} (t \to ch^0) \nonumber \\ &
 \sim \frac{g^{2}m_{c}^{2}}{4m_{W}^{2}}\tan^{2}\beta\left|\xi\right|^{2}\left[m_{h^{0}}^{2}-m_{t}^{2}\right] \label{amp2htc}~,
\end{align}
from which we get:
\begin{align}
\Gamma(h & \to\bar{t}c+\bar{c}t)=2\times\Gamma\left(h^{0}\to\bar{t}c\right) \nonumber \\ & =2N_{c}\lambda^{\frac{1}{2}}\left(1,\frac{m_{c}^{2}}{m_{h^{0}}^{2}},\frac{m_{t}^{2}}{m_{h^{0}}^{2}}\right)
\cdot\frac{\underset{pol}{\sum}\overline{\left|\mathcal{M}\right|^{2}}(h^0 \to tc)}{16\pi m_{h^{0}}}\nonumber \\
& \sim\frac{N_{c}\left|\xi\right|^{2}g^{2}m_{h^{0}}m_{c}^{2}}{32\pi m_{W}^{2}}\tan^{2}\beta\left(1-\frac{m_{t}^{2}}{m_{h^{0}}^{2}}\right)^{2},
 \end{align}
where $N_c=3$ is a color factor, and  $\underset{pol}{\sum}\overline{\left|\mathcal{M}\right|^{2}}=\underset{pol}{\sum}\left|\mathcal{M}\right|^{2}$, since the initial state is a scalar field.

To get an estimate of the $BR(h^0 \to \bar{t}c)$ we need the total width of $h^0$.
For $\alpha=\beta$ and assuming also that $m_{h^{0}}<2m_{A^{0}},2m_{H^{+}}$,
the total decay width of $h^{0}$ is mainly comprised of fermion decays,
since the couplings $W^{+}W^{-}h^{0}$, $Z^{0}Z^{0}h^{0}$ and $H^{0}H^{0}h^{0}$
are all $\propto\sin\left(\beta-\alpha\right)$ (see table \ref{tab:vvh feyn rules}
and App. \ref{app:higgs width formulae}). Thus, below the $t\bar{t}$ threshold
(at about $m_{h^0} \lsim 340$ GeV) the decay $h^0 \to b\bar{b}$ dominates, with
(see App. \ref{app:higgs width formulae}):
\begin{align}
\Gamma\left(h^{0}\to\bar{b}b\right) & \sim\frac{N_{c}g^{2}m_{b}^{2}m_{h^{0}}}{32\pi m_{W}^{2}}\tan^{2}\beta~.
\end{align}

In this case, we obtain:
\begin{align}
BR\left(h^{0}\to\bar{t}c+\bar{c}t\right) & \sim\left|\xi\right|^{2}\frac{m_{c}^{2}}{m_{b}^{2}}\left(1-\frac{m_{t}^{2}}{m_{h^{0}}^{2}}\right)^{2},
\end{align}

which for e.g., $\left|\xi\right|=0.8$ and $m_{h^{0}}=300$ GeV, amounts to
$BR\left(h^{0}\to\bar{t}c+\bar{c}t\right)\sim 0.023$.

\subsection{\label{sub:calcs 1L}The 1-loop decays $t\to cH^0$ or $H^0\to\bar{t}c$}

\begin{figure}
\includegraphics[width=245pt,keepaspectratio]{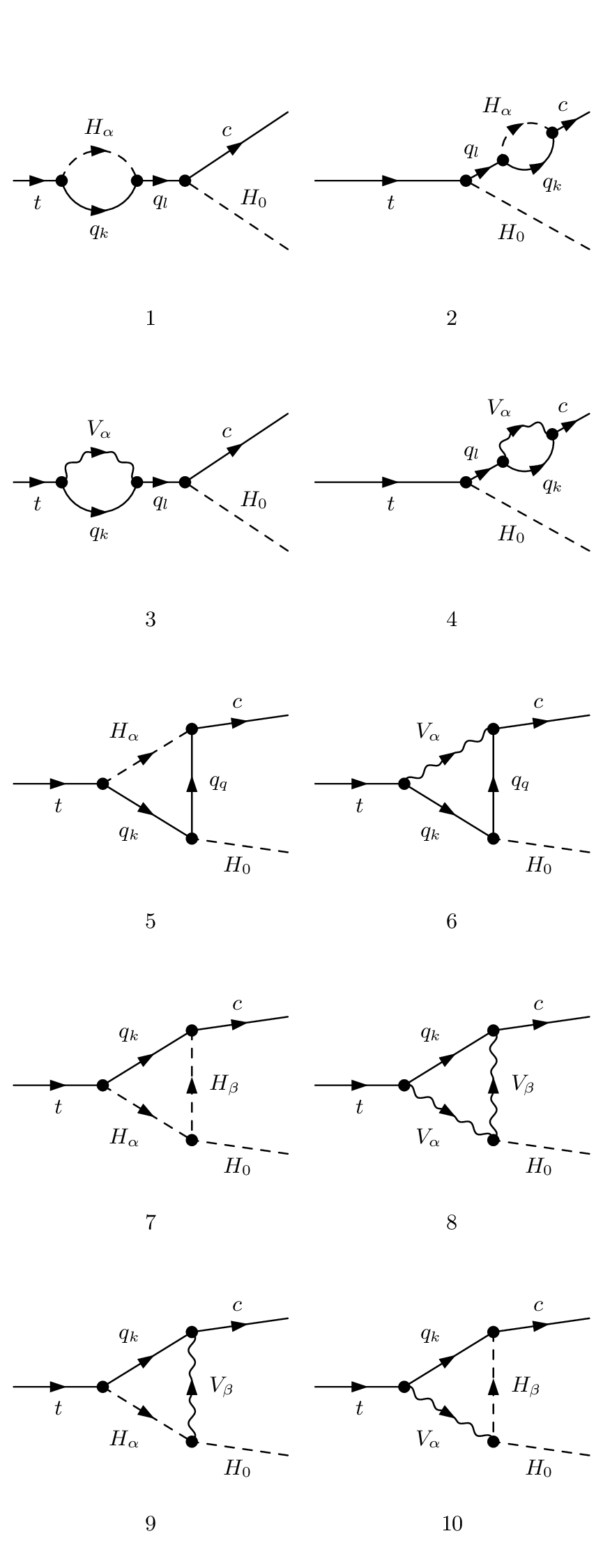}
\caption{\label{fig: 1-loop diags}1-loop Feynman diagrams for $t\to cH^{0}$}
\end{figure}

The 1-loop $t \to c H^0$ decay amplitude is composed of 10 Feynman diagrams which are shown
in Fig. \ref{fig: 1-loop diags}. The individual amplitudes corresponding to each of the 10 diagrams are given
in App. \ref{app:1L diags}. The
calculation was performed in the t'Hooft Feynman gauge and was
aimed to be as model-independent as possible, therefore assuming general vertices for the general fields $q_{i}$, $V_{\alpha}$ and $H_{\alpha}$ which stand for
a quark (up or down type), vector (gauge) fields and scalar fields, respectively (see the
calculation setup as defined by Fig. \ref{fig:vertices definition}
in App. \ref{app:1L diags}). This allowed us to easily calculate the partial width for $t \to c H^0$
(or for $H^0\to\bar{t}c$) in different
multi-Higgs models, by inserting the appropriate vertices and fields.

The 1-loop integrals were evaluated numerically with FORTRAN (f77) using
the FF package \cite{ff vanold}. The calculations were done using the Passarino-Veltman reduction scheme,
which expresses the integrals in terms of basic scalar n-point functions.
In particular, the vector and tensor integrals were computed using linear combinations
of the scalar functions (for explicit formulae see e.g. App. A in \cite{bejar}).
In App. \ref{app:dijcij definition} we describe
the reduction scheme used to calculate the 1-loop integrals.

%

As in the tree-level case, let us define the total amplitude as:
\begin{eqnarray}
\mathcal{M} & = & \frac{i\bar{u}_{c}}{16\pi^{2}}\left(M_{L}L+M_{R}R\right)u_{t} \label{1loopamp}~,
\end{eqnarray}

\noindent where
\begin{eqnarray}
M_{L,R} \equiv \sum_{i=1-10} M^i_{L,R}~,
\end{eqnarray}

\noindent and $M^i_{L,R}$ are parts of the amplitude corresponding to diagram $i$ which are
given in App. \ref{app:1L diags}.

Using Eq.~\ref{1loopamp} we can write the squared amplitude summed over polarizations as:
\begin{align}
\underset{pol}{\sum}\left|\mathcal{M}\right|^{2}= & \frac{1}{2}\frac{1}{256\pi^{4}}\left[2m_{c}m_{t}\left(M_{L}M_{R}^{*}+ M_{R}M_{L}^{*}\right)\right.\nonumber\\ & \left.+\left(m_{c}^{2}+m_{t}^{2}-m_{h}^{2}\right) \left(M_{L}M_{L}^{*}+M_{R}M_{R}^{*}\right)\right] \label{amp21loop},
\end{align}

\noindent from which we obtain:
\begin{equation}
\Gamma=4\pi\cdot\lambda^{\frac{1}{2}}\left(1,\frac{m_{c}^{2}}{m_{t}^{2}},
\frac{m_{H}^{2}}{m_{t}^{2}}\right)\cdot\frac{\underset{pol}{\sum}\overline{\left|\mathcal{M}\right|^{2}}}{64\pi^{2}m_{t}},
\end{equation}
where: $\underset{pol}{\sum}\overline{\left|\mathcal{M}\right|^{2}}=\frac{1}{2}
\underset{pol}{\sum}\left|\mathcal{M}\right|^{2}$.

Finally, the BR for the decay $t \to c H^0$ is:

\begin{equation}
BR\left(t\to cH^0\right) \sim \frac{\Gamma\left(t\to cH^0\right)}{\Gamma\left(t\to bW^+\right)}\mbox{ ,}
\end{equation}

where $\Gamma(t \to b W^+)$ is given in Eq.~\ref{tbW}.
As before, if $m_{H^{+}}<m_t$, then the top can also have an appreciable BR to
$t\to H^{+}b$ which must be taken into account for the total top width.

For the case $m_{H^0} > m_t$ we again consider the reversed 1-loop decay
$H^0 \to\bar{t}c+\bar{c}t$, where

\begin{align}
&\Gamma\left(H^0\to\bar{t}c+\bar{c}t\right)=2\times\Gamma\left(H^0\to\bar{t}c\right)\nonumber\\ &=2N_{c}\lambda^{\frac{1}{2}}\left(1,\frac{m_{c}^{2}}{m_{H^0}^{2}},\frac{m_{t}^{2}}{m_{H^0}^{2}}
\right)\cdot\frac{\underset{pol}{\sum}\overline{\left|\mathcal{M}\right|^{2}}(H^0\to tc)}{16\pi m_{H^0}},
\end{align}
where, as in the tree-level case, 
$\underset{pol}{\sum}\overline{\left|\mathcal{M}\right|^{2}}(H^0\to \bar{t}c) =\underset{pol}{\sum}\left|\mathcal{M}\right|^{2}(H^0\to \bar{t}c)
=\underset{pol}{\sum}\left|\mathcal{M}\right|^{2}(t\to cH^0)$.

Here also, in order to obtain the $BR\left(H^0\to\bar{t}c+\bar{c}t\right)$
we need to know the total width of $H^0$. We will
include the leading-order contributions to the Higgs width from $H^0\to\bar{q}q$,
$H^0\to VV$, $H^0\to\mbox{2 scalars}$ and $H^0\to\mbox{vector+scalar}$
\cite{djouadi II}. The last channel to vector+scalar can be important in some regions of the
parameter space such as low $\tan\beta$. The formulae used for the
calculation of the total $H^0$-width are given in App. \ref{app:higgs width formulae},
along with a plot (Fig. \ref{fig:higgs width SM}) of the SM Higgs width in the leading order approximation
and compared with higher order predictions (recall that for our choice of 
$\alpha=\beta$, $H^0$ behaves as
the SM-Higgs).

\section{\label{sec:Results}Numerical Results}

Before presenting our numerical results we note that we have performed several
checks to validate our calculation:

\begin{enumerate}
\item We have successfully reproduced the results for $BR(t\to ch)$
obtained in \cite{t-ch SM} for the SM case and in \cite{bejar} for the type II
2HDM case.
\item We have successfully reproduced the results for $BR\left(h\to\bar{t}c\right)$
obtained in \cite{h-tc SM arhrib} for the SM case and in \cite{arhrib}
for the type II 2HDM case. However, we were
not able to reproduce the values for $BR\left(h\to\bar{t}c\right)$ reported
in \cite{bejar}, as was also stated in \cite{arhrib}.
\item We have verified both analytically and numerically in the FORTRAN code, the cancellation of the
UV divergences which appear in some of the individual 1-loop amplitudes.
\end{enumerate}

Let us now present our results for the 1-loop
decays $BR\left(t\to cH^{0}\right)$ and $BR\left(H^{0}\to\bar{t}c\right)$
in the T2HDM. We have taken the following set of assumptions/values on the relevant parameter
space of the T2HDM:

\begin{itemize}
\item Set $\alpha=\beta$ for the reasons explained above.
\item Set $\left|\xi\right| \sim 0.8$, as in \cite{soni 34 best}.
\item The other parameters of the T2HDM are set to their
best-fit (central) value in \eqref{eq:bounds} unless stated otherwise.
\item For the analysis of the decay $t\to cH^{0}$ we set $m_{H^{0}}=91$ GeV, which
is the central (best fitted) value
of the SM Higgs mass to EW precision data
\cite{PDG}. Recall that, in our setup, $H^{0}$ has couplings identical
to the SM Higgs and we therefore expect the phenomenology of $H^0$ to
roughly follow that of the SM's Higgs.
\item For the total top decay width we take $\Gamma\left(t\to W^{+}b\right)=1.55$ GeV.
\item For the process $H^{0}\to\bar{t}c$ we arbitrarily choose
$m_{H^{0}}=300$ GeV.
\item We set $m_{A^{0}}\sim 1$ TeV to enhance the triple-scalar
coupling, which is roughly $\propto m_{A^{0}}^{2}$ (see App. \ref{app:Feynman-rules}).
\item Other values used for the calculations were \cite{PDG}:
$m_{t}=172.5$ GeV (pole mass), $m_{c}=\overline{m}_{c}\left(\overline{m}_{c}\right)=1.24$ GeV,
$m_{b}=\overline{m}_{b}\left(\overline{m}_{b}\right)=4.20$ GeV
($m_{c}$ and $m_{b}$ are in the $\overline{MS}$ renormalization
scheme), $m_{W}=80.40$ GeV, $m_{Z}=91.188$ GeV, $\cos\theta_{W}=m_{W}/m_{Z}$,
$\alpha\left(m_{z}\right)=1/128$. The mass 
values used are without running the energy scale, even though
the BR's were found to be sensitive to both $m_{c}$ and $m_{b}$ (our results are not sensitive to $m_{s}$).
For example,
the value $BR(t\to c H^0) = 5.99\times10^{-5}$ quoted
in the upper right corner of Table \ref{tab:t-ch results compare} would change
to $BR(t\to c H^0) = 1.28\times10^{-5}$ had we used
$m_{b}(m_{Z})\sim3$ GeV and $m_{c}(m_{Z})\sim 0.7$ GeV 
\cite{running masses 2000}.
In addition see  below for an example with explicit scale dependence. 
\end{itemize}

\subsection{\label{sub:res t-hc}The 1-loop decay $t \to c H^{0}$}

\begin{figure}
\begin{tabular}{l}
\vspace{-0.75cm}
\tabularnewline
\includegraphics[width=8cm,keepaspectratio]{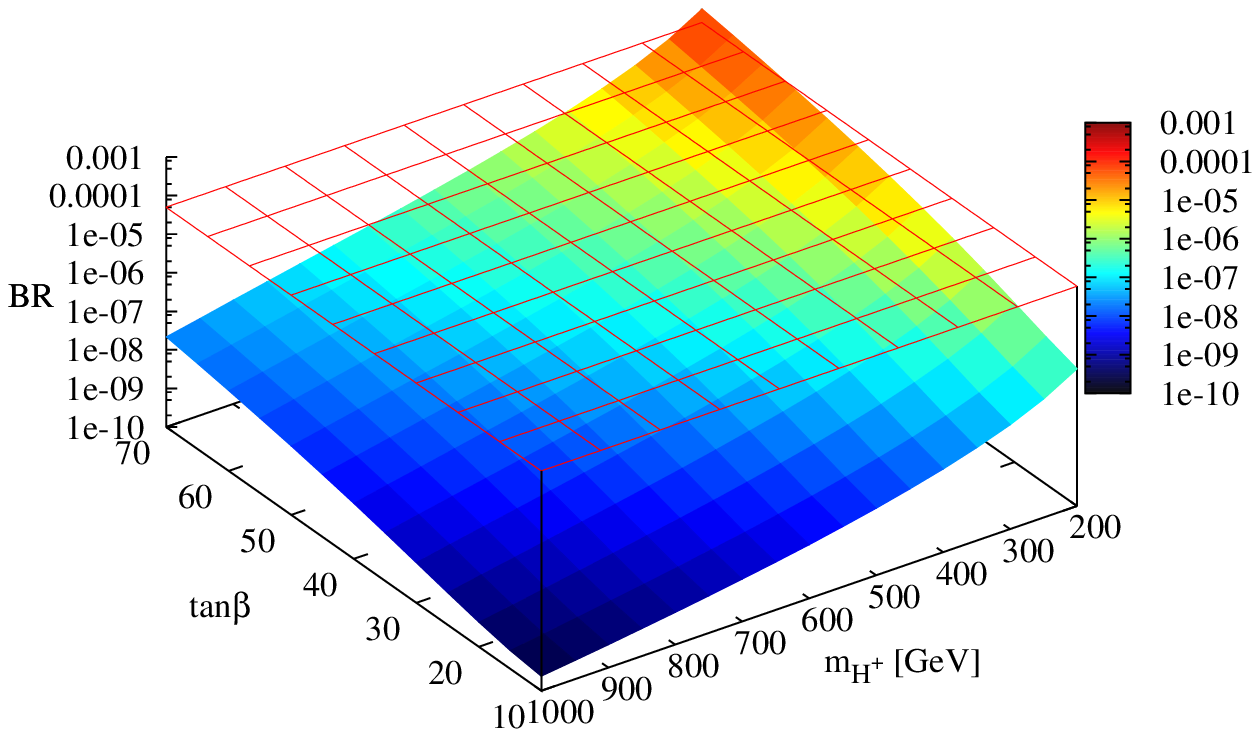}
\vspace{-0.75cm}
\tabularnewline
(a)
\tabularnewline
\vspace{-1cm}
\tabularnewline
\includegraphics[width=120pt,keepaspectratio]{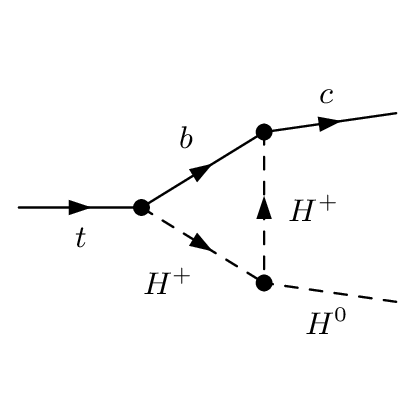}
\vspace{-1cm}
\tabularnewline
(b)
\tabularnewline
\end{tabular}
\caption{\label{fig:t-ch 3D mH+:tanb}(a) 3D plot of $BR(t\to cH^{0})$
in the $m_{H^{+}}-\tan\beta$ plane in the T2HDM, and (b) the dominant
diagram. We set $m_{h^{0}}=1000$ GeV and $m_{A^{0}}=1200$ GeV.
The color scale represents the BR: the blue represents the lowest
BR and red the highest.}
\end{figure}

\begin{figure}
\begin{tabular}{l}
\vspace{-0.75cm}
\tabularnewline
\includegraphics[width=8cm,keepaspectratio]{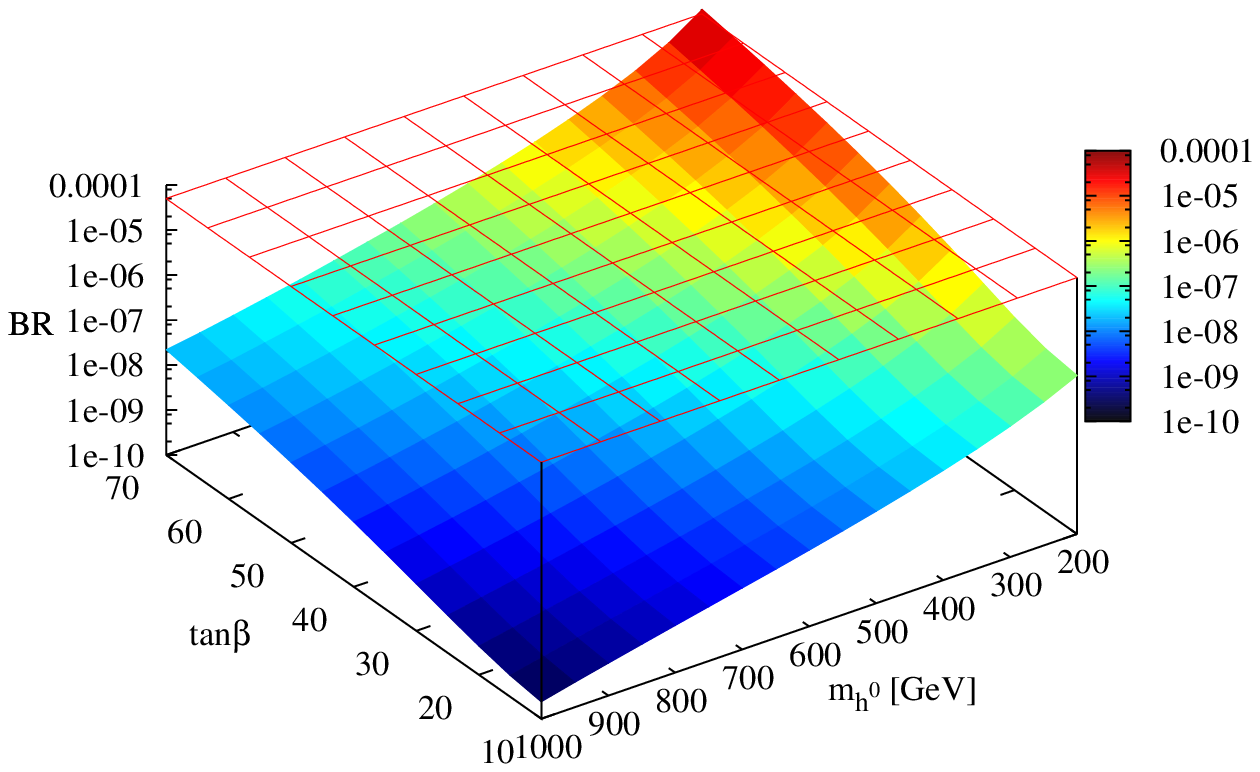}
\vspace{-0.75cm}
\tabularnewline
(a)
\tabularnewline
\vspace{-1cm}
\tabularnewline
\includegraphics[width=120pt,keepaspectratio]{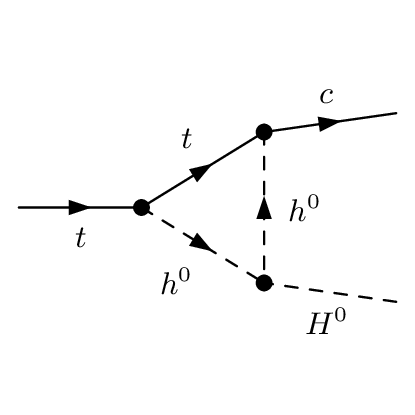}
\vspace{-1.25cm}
\tabularnewline
(b)
\tabularnewline
\end{tabular}
\caption{\label{fig:t-ch 3D tanb:mh0}(a) 3D plot of $BR(t\to cH^{0})$
in the $m_{h^{0}}-\tan\beta$ plane in the T2HDM, and (b) the dominant
diagram. We set $m_{H^{+}}=1000$ GeV and $m_{A^{0}}=1200$ GeV. 
See also caption to Fig.~\ref{fig:t-ch 3D mH+:tanb}.}
\end{figure}

In Figs.~\ref{fig:t-ch 3D mH+:tanb}a and \ref{fig:t-ch 3D tanb:mh0}a we give a 3D plot
of $BR\left(t\to cH^{0}\right)$ in the $m_{H^{+}}-\tan\beta$ and $m_{h^0}-\tan\beta$ planes, respectively.
The flat grid in Figs.~\ref{fig:t-ch 3D mH+:tanb}a and \ref{fig:t-ch 3D tanb:mh0}a
represents the LHC detection limit which is $\sim 5\cdot10^{-5}$, so that the colored
surface above the grid is the region of the parameter space of the T2HDM which
has a BR potentially within the sensitivity of the LHC.

The choice $m_{h^{0}}=1000$ GeV made in Fig. \ref{fig:t-ch 3D mH+:tanb}a
suppresses the diagrams with $h^{0}$ in   
the loop and, thus, better explores the charged Higgs sector.
As expected the BR rises with $\tan\beta$ and is highest when $m_{H^{+}}$
is lowest. The dominant Feynman diagram in this case is the one
depicted in Fig.~\ref{fig:t-ch 3D mH+:tanb}b with two $H^{+}$ and a $b$-quark in the loop.
This diagram receives an enhancement from the 3-scalar vertex $H^0H^+H^-$, as noted above.

The choice $m_{H^{+}}=1000$ GeV made in Fig.~\ref{fig:t-ch 3D tanb:mh0}a
suppresses the diagrams with $H^\pm$ in
the loop and, thus, is more sensitive to the neutral Higgs sector.
In this case also, the BR rises with $\tan\beta$ and drops as $m_{h^{0}}$ is
increased. The dominant diagram in this
case is the one which has two $h^{0}$ scalars
in the loop. This diagram receives
an enhancement from the 3-scalar vertex $H^0h^0h^0$ when $m_{A^{0}}$ is large, as
mentioned above.

The two limits $m_{h^{0}}=1000$ GeV and $m_{H^{+}}=1000$ GeV have similar
consequences, yet the $BR\left(t\to cH^{0}\right)$ is higher when $m_{H^{+}}< m_{h^{0}}$
in which case the charged Higgs loop-exchange is the dominant source for
the enhanced $BR\left(t\to cH^{0}\right)$.
This is a distinctive property of the T2HDM, since, in this model, the charged
Higgs coupling $H^{+}\bar{b}c$ is enhanced by $V_{tb}/V_{cb}$
compared to other 2HDM's such as the type I and type II 2HDM.

\begin{figure}
\vspace{-1cm}
\begin{centering}
\includegraphics[width=8cm]{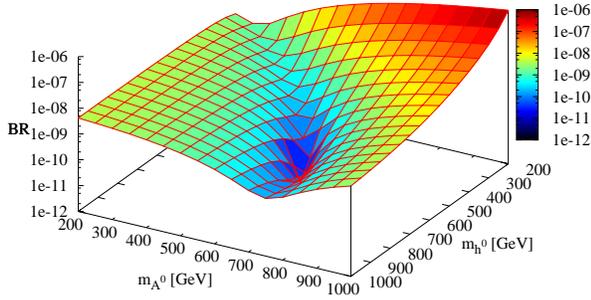}
\par\end{centering}
\vspace{-0.5cm}
\caption{\label{fig:t-ch 3D mA0:mh0}3D plot of $BR(t\to cH^{0})$
in the $m_{A^{0}}-m_{h^{0}}$ plane in the T2HDM. We set $m_{H^{+}}=660$ GeV
and $\tan\beta=28$.}
\end{figure}

In Fig. \ref{fig:t-ch 3D mA0:mh0} we plot the $BR\left(t\to cH^{0}\right)$
in the $m_{A^{0}}-m_{h^{0}}$ plane, where all other parameters are given the
central values of Eq.~\eqref{eq:bounds}.
We see that the BR is highest for a large $m_{A^{0}}$ and
a small $m_{h^{0}}$ where the diagram with the two $h^{0}$ in the loop
dominates. The dip in the middle of the surface is due to
cancellations in the $H^{0} H^{+} H^{+}$ vertex.

\begin{figure}
\begin{centering}
\includegraphics[width=8cm,keepaspectratio]{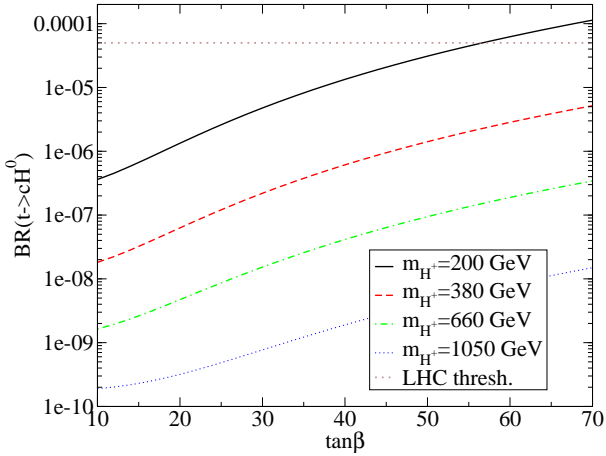}
\par\end{centering}
\caption{\label{fig:t-ch tanb}The $BR(t\to cH^{0})$ as
a function of $\tan\beta$ at various $m_{H^{+}}$ in the T2HDM. We
set $m_{h^{0}}=1000$ GeV and $m_{A^{0}}=1200$ GeV.
``LHC thresh.'' stands for the limit of the LHC sensitivity at
100 $fb^{-1}$.}
\end{figure}

\begin{figure}
\begin{centering}
\includegraphics[width=8cm,keepaspectratio]{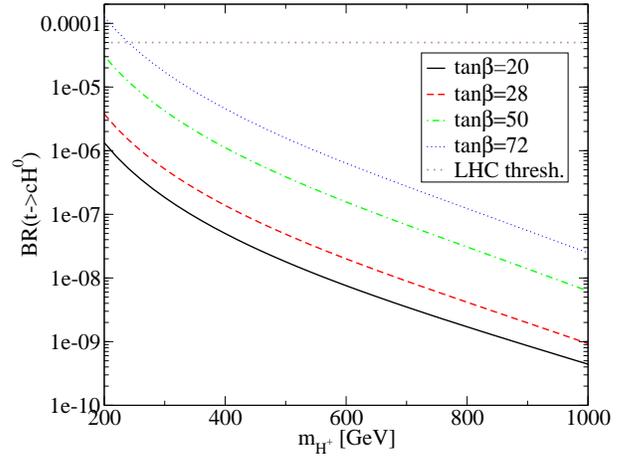}
\par\end{centering}
\caption{\label{fig:t-ch mH+}The $BR(t\to cH^{0})$ as
a function of $m_{H^{+}}$ at various $\tan\beta$ in the T2HDM. We
set $m_{h^{0}}=1000$ GeV and $m_{A^{0}}=1200$ GeV.
{}``LHC thresh.'' stands for the limit of the LHC sensitivity at
100 $fb^{-1}$.}
\end{figure}

To better illustrate the dependence of the $BR\left(t\to cH^{0}\right)$ on $\tan\beta$ and $m_{H^{+}}$
we give in Figs. \ref{fig:t-ch tanb} and \ref{fig:t-ch mH+} 2D plots
of the $BR\left(t\to cH^{0}\right)$ as a function of $\tan\beta$ and $m_{H^{+}}$, respectively,
using the same parameter set as in Fig. \ref{fig:t-ch 3D mH+:tanb}.

Finally, in order to demonstrate the difference in the
$BR\left(t\to cH^{0}\right)$ expected within
the T2HDM, the 2HDM-II, and the SM , we list in table
\ref{tab:t-ch results compare} the $BR\left(t\to cH^{0}\right)$
values within these 3 different models, for 4 different points in the
relevant parameter space. Note that in the SM the
1-loop $BR\left(t\to cH^{0}\right)$ depends only on the SM's Higgs mass
which we also set to $m_{H^{0}}=91$ GeV.
Also, recall that the type II 2HDM Yukawa
potential is similar to that of the MSSM and that it has no tree-level
FCNC.

\begin{table*}
\begin{tabular}{|c|c|c|c|}
\hline
\vphantom{$a^b$}
parameters&
SM&
2HDM-II&
T2HDM\tabularnewline
\hline
$m_{h^{0}}=800$, $m_{A^{0}}=1000$, $\tan\beta=72$, $m_{H^{+}}=200$&
$\vphantom{A^{A^A}} 6.03\times10^{-14}$&
$4.25\times10^{-5}$&
$5.99\times10^{-5}$\tabularnewline
\hline
$m_{h^{0}}=800$, $m_{A^{0}}=1000$, $\tan\beta=72$, $m_{H^{+}}=380$&
$\vphantom{A^{A^A}} 6.03\times10^{-14}$&
$1.79\times10^{-6}$&
$2.57\times10^{-6}$\tabularnewline
\hline
$m_{h^{0}}=200$, $m_{A^{0}}=4000$, $\tan\beta=20$, $m_{H^{+}}=1050$&
$\vphantom{A^{A^A}} 6.03\times10^{-14}$&
$5.15\times10^{-8}$&
$9.39\times10^{-5}$\tabularnewline
\hline
$m_{h^{0}}=200$, $m_{A^{0}}=1000$, $\tan\beta=20$, $m_{H^{+}}=1050$&
$\vphantom{A^{A^A}} 6.03\times10^{-14}$&
$3.34\times10^{-12}$&
$3.14\times10^{-7}$\tabularnewline
\hline
\end{tabular}
\caption{\label{tab:t-ch results compare}Comparison of the 
$BR(t\to cH^{0})$
within the T2HDM, the 2HDM-II, and the SM. Masses are in units of
GeV.}
\end{table*}

The first two rows in table
\ref{tab:t-ch results compare} illustrate the impact of the charged sector, by
setting a high $m_{h^{0}}$ and a much smaller $m_{H^{+}}$ (note that the value $m_{H^{+}}=200$ GeV
is outside the $1\sigma$ bounds). In this case, the $BR\left(t\to cH^{0}\right)$ in the
T2HDM is not as enhanced
as expected relative to the 2HDM-II, where it is a bit higher in the T2HDM.
Recall that we expected the diagram
with the $H^+$-$b$ quark in the loop to be particularly enhanced in the T2HDM due
to the enhanced $H^{+}\bar{c}b$ interaction in this model.
The amplitude of this diagram is (see App. \ref{app:dijcij definition}):
\begin{align}
M_{7} & =\frac{-i\bar{u}_{c}}{16\pi^{2}}g_{H^{+}H^{+}h}^{3h} \left[m_{b}C_{0}\left(A_{cb}^{H^{+}}B_{tb}^{H^{+}*}L+B_{cb}^{H^{+}}A_{tb}^{H^{+}*}R\right)\right.\nonumber\\
&-m_{c}C_{12}\left(B_{cb}^{H^{+}}B_{tb}^{H^{+}*}L+A_{cb}^{H^{+}}A_{tb}^{H^{+}*}R\right)\nonumber\\
&\left.+m_{t}\left(-C_{11}+C_{12}\right)\left(A_{cb}^{H^{+}}A_{tb}^{H^{+}*}L+B_{cb}^{H^{+}}B_{tb}^{H^{+}*}R\right)\right]u_{t},
\label{eq:amp7 H+H+b}
\end{align}
where $C_{ij}$ are the Passarino-Veltman scalar functions.
The term $\propto m_{b}A_{cb}^{H^{+}}B_{tb}^{H^{+}*}$ (multiplied by the left projection operator),
which is sub-leading in the type II 2HDM,
is enhanced in the T2HDM and dominates the other terms,
being $\propto\xi^{*}m_{c}m_{b}^{2}\tan^{2}\beta V_{tb}V_{tb}^{*}$.
On the other hand, in the 2HDM of type II it is the term
$\propto m_{t}B_{cb}^{H^{+}}B_{tb}^{H^{+}*} \sim m_{t}m_{b}^{2}\tan^{2}\beta V_{cb}V_{tb}^{*}$
which dominates for a large $\tan\beta\gtrsim10$.
Therefore, we see that the different leading terms in 
the T2HDM and the type II 2HDM
are roughly of the same order of magnitude since $m_t \cdot V_{cb} \sim m_c$,
and therefore the enhancement in the T2HDM is not as significant as expected.

\begin{figure}[htb]
\begin{centering}
\includegraphics[width=8cm,keepaspectratio]{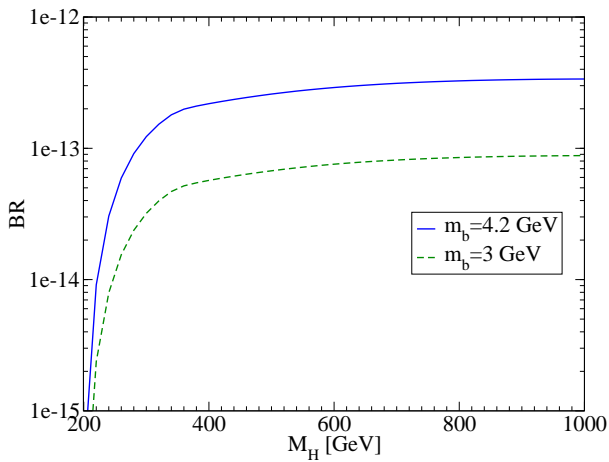}
\par\end{centering}
\caption{\label{fig:h-tc SM} The SM value for the 
$BR(H^{0}\to\bar{t}c+\bar{c}t)$
as a function of the Higgs mass, for $\overline{m_{b}}(\overline{m_{b}})=4.2$ GeV
and for $\overline{m_{b}}(\overline{m_{Z}})=3$ GeV \cite{PDG}.
The BR is not sensitive to $m_{c}$.}
\end{figure}

\begin{figure}[htb]
\begin{tabular}{l}
\vspace{-1cm}
\tabularnewline
\includegraphics[width=8cm,keepaspectratio]{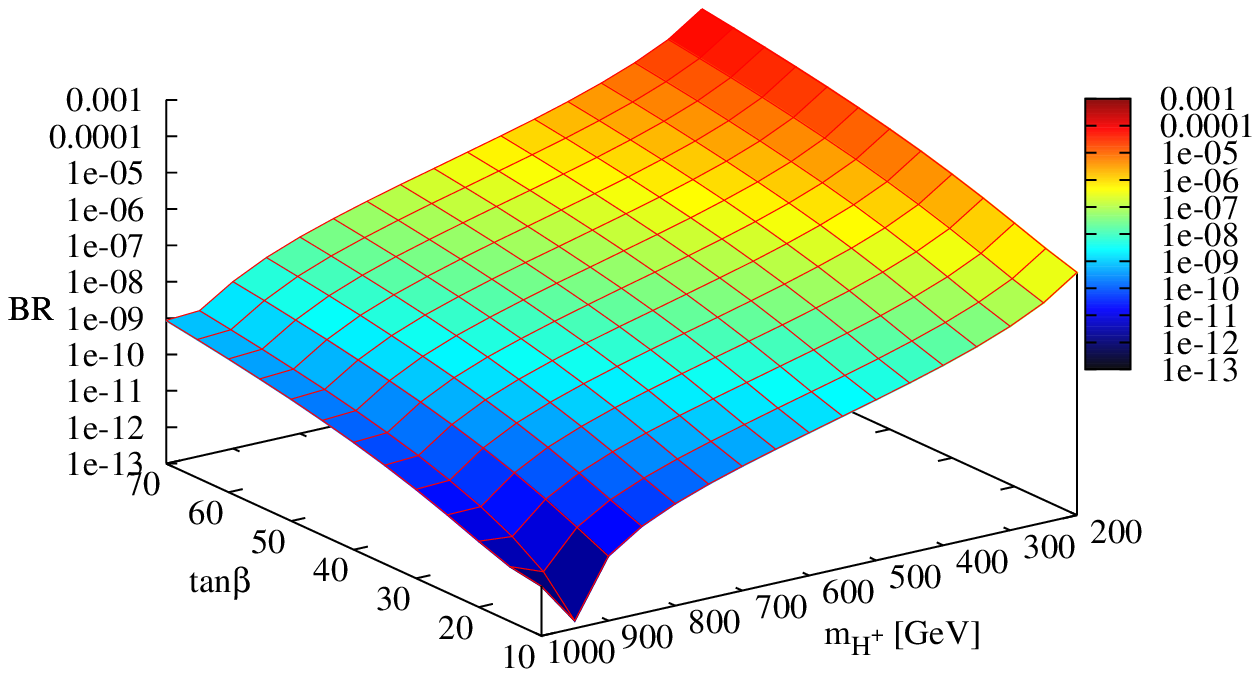}
\vspace{-1cm}
\tabularnewline
(a)
\tabularnewline
\vspace{-1cm}
\tabularnewline
\includegraphics[width=120pt,keepaspectratio]{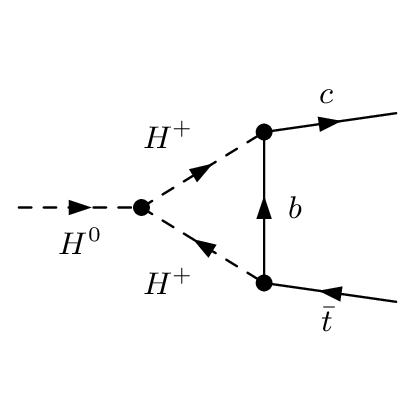}
\vspace{-1.25cm}
\tabularnewline
(b)
\tabularnewline
\end{tabular}
\caption{\label{fig:h-tc 3D tanb:mhp}(a) 3D plot of $BR(H^{0}\to\bar{t}c+\bar{c}t)$
in the $m_{H^{+}}-\tan\beta$ plane in the T2HDM, and (b) the
dominant diagram. We set $m_{h^{0}}=1000$ GeV and $m_{A^{0}}=1000$ GeV.}
\end{figure}

\begin{figure}[htb]
\begin{centering}
\includegraphics[width=8cm,keepaspectratio]{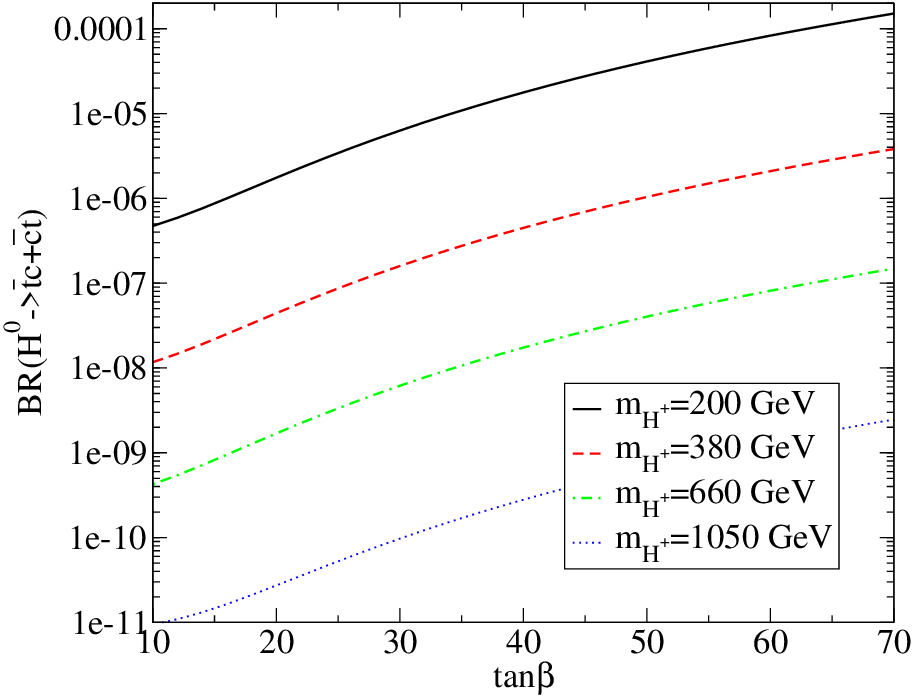}
\par\end{centering}
\caption{\label{fig:h-tc tanb var-mhp}The 
$BR(H^{0}\to\bar{t}c+\bar{c}t)$
as a function of $\tan\beta$ at different $m_{H^{+}}$ in the T2HDM.
We set $m_{h^{0}}=1000$ GeV and $m_{A^{0}}=1000$ GeV.}
\end{figure}

\begin{figure}[htb]
\begin{centering}
\includegraphics[width=8cm,keepaspectratio]{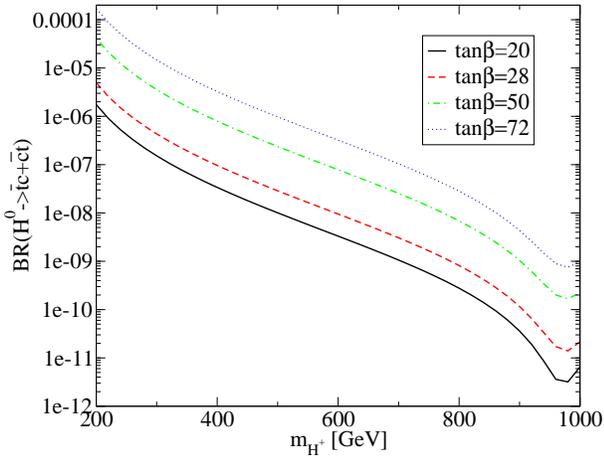}
\par\end{centering}
\caption{\label{fig:h-tc mhp var-tanb}The $BR(H^{0}\to\bar{t}c+\bar{c}t)$
as a function of $m_{H^{+}}$ at different $\tan\beta$ in the T2HDM.
We set $m_{h^{0}}=1000$ GeV and $m_{A^{0}}=1000$ GeV.}
\vspace{-0.5cm}
\end{figure}

\begin{figure}[htb]
\begin{tabular}{l}
\vspace{-1cm}
\tabularnewline
\includegraphics[width=8cm,keepaspectratio]{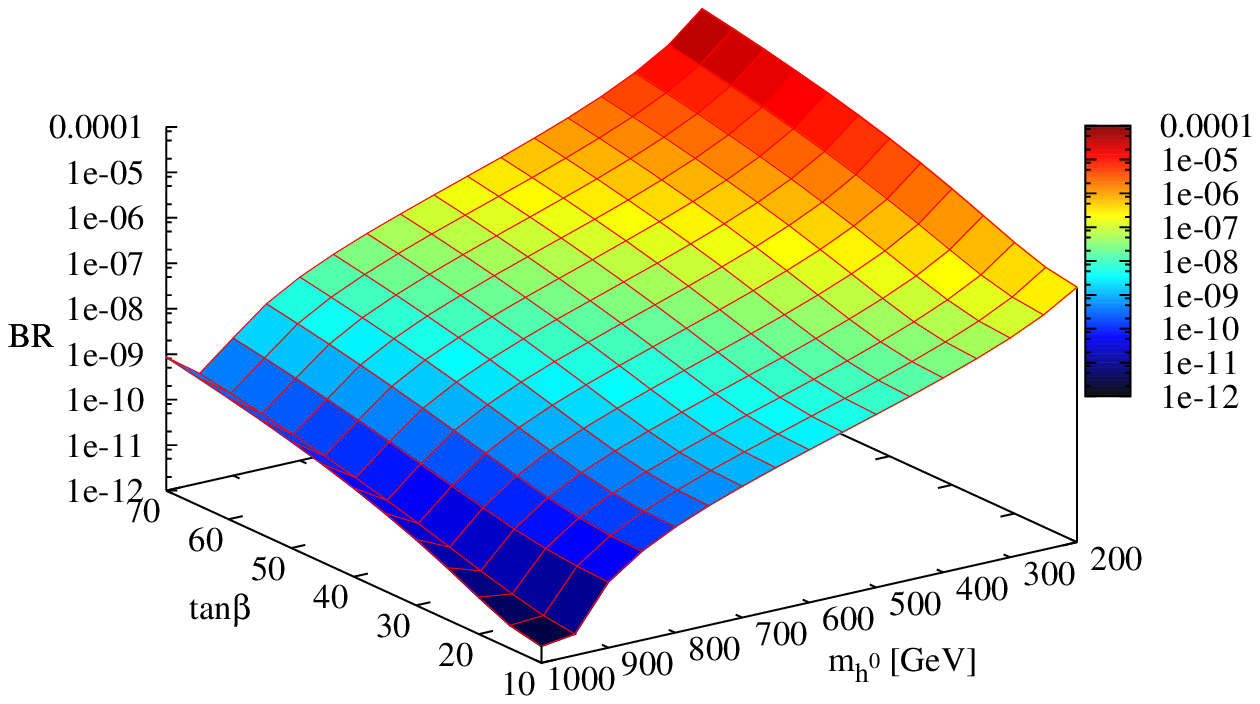}
\vspace{-1cm}
\tabularnewline
(a)
\tabularnewline
\vspace{-1.25cm}
\tabularnewline
\includegraphics[width=120pt,keepaspectratio]{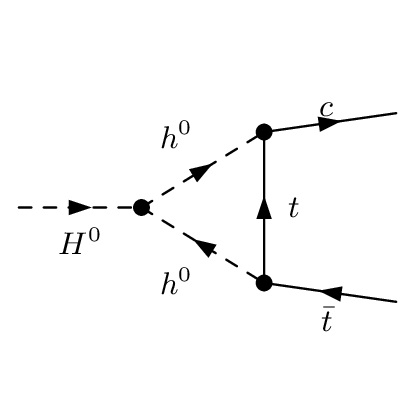}
\vspace{-1cm}
\tabularnewline
\vspace{-0.5cm}
\tabularnewline
(b)
\end{tabular}
\caption{\label{fig:h-tc 3D tanb:mh0}(a) 3D plot of 
$BR(H^{0}\to\bar{t}c+\bar{c}t)$
in the $m_{h^{0}}-\tan\beta$ plane in the T2HDM, and (b) the
dominant diagram. We set $m_{H^{+}}=1000$ GeV and $m_{A^{0}}=1000$ GeV.}
\end{figure}

\begin{figure}[htb]
\begin{centering}
\includegraphics[width=8cm,keepaspectratio]{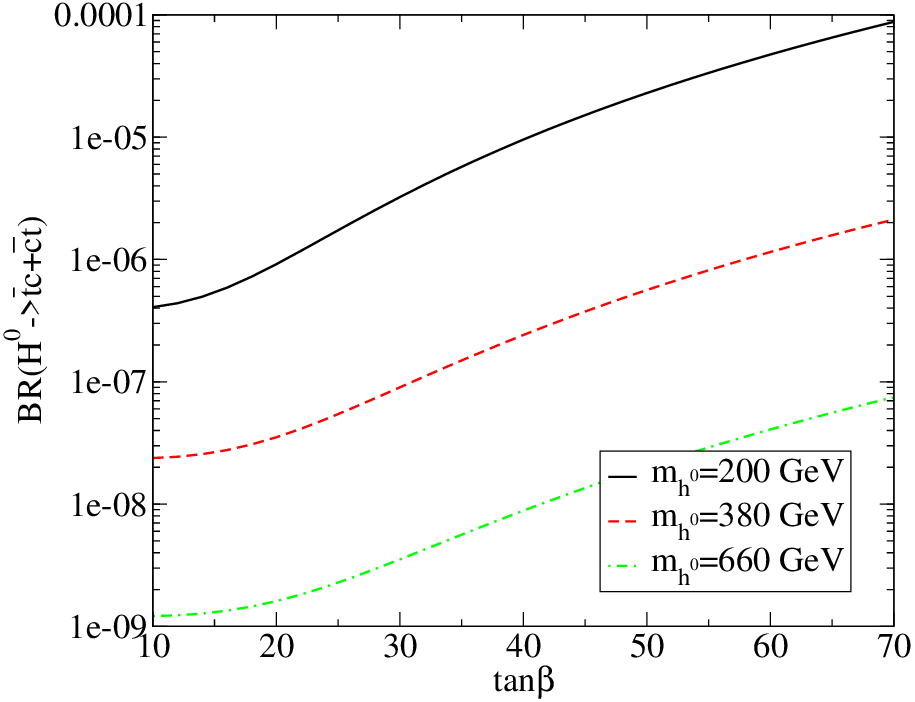}
\par\end{centering}
\vspace{-0.5cm}
\caption{\label{fig:h-tc tanb var-mh0}The $BR(H^{0}\to\bar{t}c+\bar{c}t)$
as a function of $\tan\beta$ at different $m_{h^{0}}$ in the T2HDM.
We set $m_{H^{+}}=1000$ GeV and $m_{A^{0}}=1000$ GeV.}
\end{figure}

\begin{table*}[htb]
\begin{tabular}{|c|c|c|c|}
\hline
\vphantom{$a^b$}
parameters&
SM&
2HDM-II&
T2HDM\tabularnewline
\hline
$m_{h^{0}}=800$, $m_{A^{0}}=1000$, $\tan\beta=72$, $m_{H^{+}}=200$&
$\vphantom{A^{A^A}} 1.23\times10^{-13}$&
$1.26\times10^{-4}$&
$1.70\times10^{-4}$\tabularnewline
\hline
$m_{h^{0}}=800$, $m_{A^{0}}=1000$, $\tan\beta=72$, $m_{H^{+}}=380$&
$\vphantom{A^{A^A}} 1.23\times10^{-13}$&
$3.09\times10^{-6}$&
$4.45\times10^{-6}$\tabularnewline
\hline
$m_{h^{0}}=200$, $m_{A^{0}}=4000$, $\tan\beta=20$, $m_{H^{+}}=1050$&
$\vphantom{A^{A^A}} 1.23\times10^{-13}$&
$8.69\times10^{-8}$&
$2.90\times10^{-4}$\tabularnewline
\hline
$m_{h^{0}}=200$, $m_{A^{0}}=1000$, $\tan\beta=20$, $m_{H^{+}}=1050$&
$\vphantom{A^{A^A}} 1.23\times10^{-13}$&
$8.99\times10^{-12}$&
$9.11\times10^{-7}$\tabularnewline
\hline
\end{tabular}
\caption{\label{tab:h-tc results compare}Comparison of 
$BR(H^{0}\to\bar{t}c+\bar{c}t)$
between the T2HDM, the 2HDM-II, and the SM. Masses are in units of
GeV. We set $m_{H^{0}}=300$, $\alpha=\beta$, and other parameters
to their best-fit value of \eqref{eq:bounds}.}
\end{table*}

In the last two rows of table
\ref{tab:t-ch results compare} we set a high $m_{H^{+}} \sim 1$ TeV,
thus exploring the impact of an EW-scale neutral
Higgs sector. Evidently, in this case the $BR\left(t\to cH^{0}\right)$ is much
larger in the T2HDM than in the 2HDM of type II. This is in fact expected
since the type II 2HDM does not have any tree-level FCNC.

\subsection{\label{sub:res h-tc}The 1-loop decay $H^{0}\to\bar{t}c$}

In the results that follow, we assume that the Higgs decays that enter its total width (when kinematically
allowed) are: $H^{0}\to\bar{q}q$, $H^{0}\to VV$,
$H^{0}\to h_{i}h_{j}$ and $H^{0}\to V_{i}h_{j}$. The partial widths for these decay channels are
given in App. \ref{app:higgs width formulae}.

We first plot in Fig. \ref{fig:h-tc SM} the SM value for the
$BR\left(H^{0}\to\bar{t}c+\bar{c}t\right)$, as a function of the Higgs mass, for
two b-quark masses: $\overline{m_{b}}(\overline{m_{b}})=4.2$ GeV
and $\overline{m_{b}}(\overline{m_{Z}})=3$ GeV. Our results
are in agreement with the results reported in \cite{h-tc SM arhrib}.

Next we turn to our results in the T2HDM. In Fig. \ref{fig:h-tc 3D tanb:mhp}
we give a 3D plot of $BR\left(H^{0}\to\bar{t}c+\bar{c}t\right)$
in the $m_{H^{+}}-\tan\beta$ plane and
in Figs.~\ref{fig:h-tc tanb var-mhp} and \ref{fig:h-tc mhp var-tanb}
we plot (2D) the BR as a function of $\tan\beta$ and $m_{H^{+}}$,
respectively, with the same parameters as in Fig. \ref{fig:h-tc 3D tanb:mhp}.
We again see the
same tendency as in the case of $t\to cH^{0}$, i.e., the BR rises
with $\tan\beta$ and decreases with $m_{H^{+}}$.

In Fig. \ref{fig:h-tc 3D tanb:mh0} we give a 3D plot of the $BR\left(H^{0}\to\bar{t}c+\bar{c}t\right)$
in the $m_{h^{0}}-\tan\beta$ plane, and in Fig.
\ref{fig:h-tc tanb var-mh0} we plot (2D) the BR as a function
of $\tan\beta$ with the same parameters as Fig. \ref{fig:h-tc 3D tanb:mh0}
for several values of $m_{h^{0}}$. We again see that the BR decreases with
$m_{h^{0}}$ and increases with $\tan\beta$.

Finally, in table \ref{tab:h-tc results compare} we give the $BR\left(H^{0}\to\bar{t}c+\bar{c}t\right)$
in the three different models (SM, type II 2HDM and the T2HDM) for 4 points of the
relevant parameter space. As can be seen, the behavior is similar to the reversed top decay
$t\to cH^{0}$ process, albeit the $BR\left(H^{0}\to\bar{t}c+\bar{c}t\right)$ are typically higher.

\section{Summary}

We have studied the top and neutral Higgs FCNC rare decays 
$t\to ch$ and $h\to\bar{t}c$ ($h= h^0$ or $H^0$ are the two CP-even neutral Higgs) 
within the T2HDM.
In this model the Higgs doublet with the heavier VEV ($v_2$) couples only to 
the top-quark, while the lighter Higgs doublet (i.e., with $v_1 \ll v_2$) couples to 
all other quarks. In particular, the
working assumption of the T2HDM is that $\tan\beta\equiv v_{2}/v_{1} \gg 1$, so that
the top quark receives a much larger mass than all other quarks in a natural manner. 

The Yukawa sector of the T2HDM exibits potentially enhanced
FCNC in the up-quark sector and large flavor transitions
mediated by the charged Higgs. These Yukawa interactions and the
scalar self interactions of the model were explicitly (and independently)
derived.
For example, it was shown that the $H^{+}\bar{b}c$ Yukawa
coupling, which (in this model) is enhanced by a factor of $V_{tb}/V_{cb}$ compared to the
corresponding 2HDM type II coupling, enhances the 1-loop $t\to cH^{0}$
and $H^{0}\to\bar{t}c$ decays via diagrams involving $H^{+}$ and $b$-quarks 
inside the loop.
Another potential enhancement of these 1-loop decays can come from
the FCNC $h^0\bar{t}c$ Yukawa interaction, i.e.,
via diagrams containing $h^0$ and $t$-quarks.

Without loss of generality, we have considered the region 
in parameter space in which decays involving $h^0$ occur at tree-level while 
those involving $H^0$ are 1-loop mediated. 
We then explored the parameter space of the T2HDM for the resulting decays 
and found that the BR's for the tree-level decays
$t\to ch^0$ and $h^0\to\bar{t}c$ are typically of ${\cal O}(0.01)$, while 
the BR's for the 1-loop decays $t\to cH^0$ and $H^0\to\bar{t}c$ can reach 
$10^{-5} - 10^{-4}$ in a favorable scenario - a value higher than 
the LHC detection threshold for the top-decay and above 
their expected value within the SM and the type I and II 2HDM.
Thus, even if $h^0$ decouples (i.e., too heavy), the 1-loop FCNC 
top-decay $t\to cH^0$ may still be accessible to the LHC.

\appendix

\section{\label{app:Feynman-rules}Feynman rules for two Higgs doublet models}

\begin{figure}[htb]
\includegraphics[width=250pt,keepaspectratio]{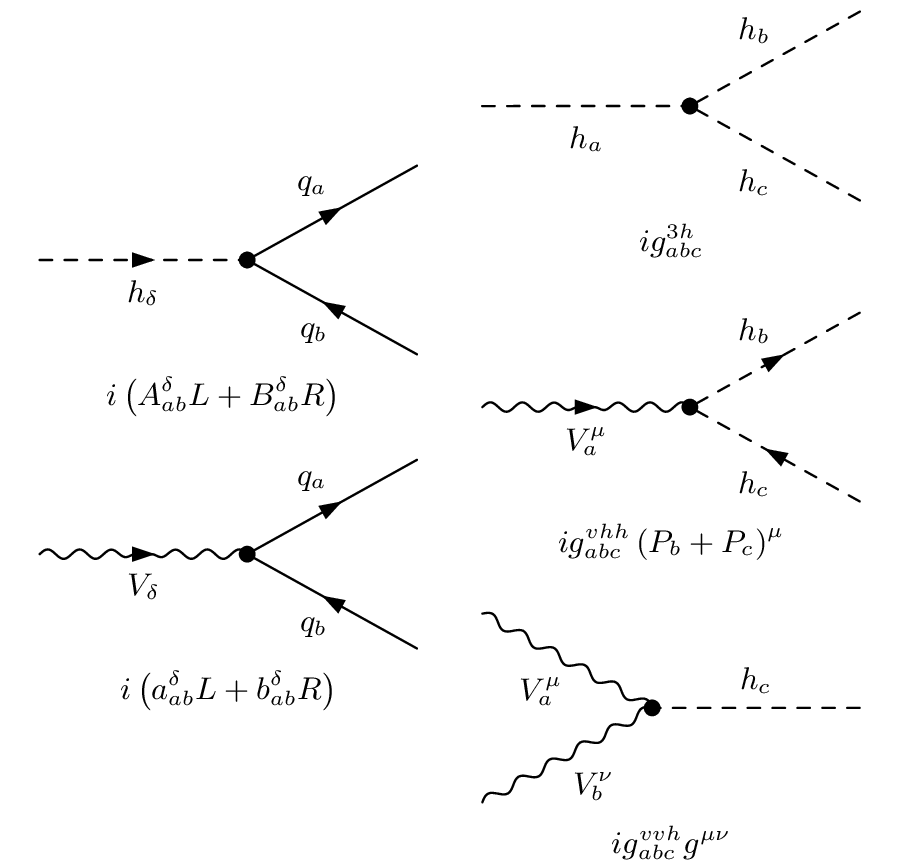}
\vspace{-0.5cm}
\caption{\label{fig:vertices definition}Feynman rules.}
\end{figure}

\begin{table}[htb]
\begin{tabular}{|c|c|}
\hline
\begin{tabular}{c}
\vspace{-1cm}
\tabularnewline
\includegraphics[width=110pt,keepaspectratio]{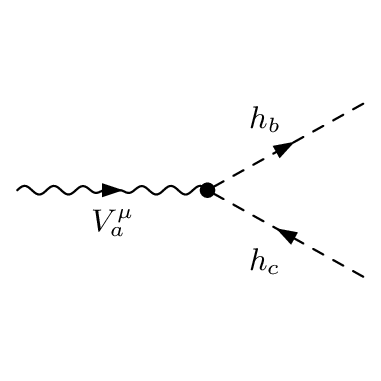}
\vspace{-1cm}
\tabularnewline
\end{tabular}&
\begin{tabular}{c}
\vspace{-0.5cm}
\tabularnewline
$=ig_{abc}^{vhh}\left(P_{b}+P_{c}\right)^{\mu}$\tabularnewline
\end{tabular}\tabularnewline
\hline
$W^{+}H^{+}H^{0}$&
$i\frac{g}{2}\sin\left(\beta-\alpha\right)\left(P_{H^{+}}+P_{H^{0}}\right)^{\mu}$\tabularnewline
\hline
$W^{+}H^{+}h^{0}$&
$-i\frac{g}{2}\cos\left(\beta-\alpha\right)\left(P_{H^{+}}+P_{h^{0}}\right)^{\mu}$\tabularnewline
\hline
$W^{+}G^{+}H^{0}$&
-$i\frac{g}{2}\cos\left(\beta-\alpha\right)\left(P_{G^{+}}+P_{H^{0}}\right)^{\mu}$\tabularnewline
\hline
$W^{+}G^{+}h^{0}$&
$-i\frac{g}{2}\sin\left(\beta-\alpha\right)\left(P_{G^{+}}+P_{h^{0}}\right)^{\mu}$\tabularnewline
\hline
$Z^{0}A^{0}H^{0}$&
$-\frac{g\sin\left(\beta-\alpha\right)}{2\cos\theta_{W}}\left(P_{A^{0}}+P_{H^{0}}\right)^{\mu}$\tabularnewline
\hline
$Z^{0}A^{0}h^{0}$&
$\frac{g\cos\left(\beta-\alpha\right)}{2\cos\theta_{W}}\left(P_{A^{0}}+P_{h^{0}}\right)^{\mu}$\tabularnewline
\hline
\end{tabular}
\caption{\label{tab:vhh feyn rules}Feynman rules for vector-scalar-scalar
interactions as in \cite{HHG}. The second
particle is outgoing.}
\end{table}

%
\begin{table}
\vspace{0.5cm}
\begin{tabular}{|c|c|}
\hline
$H^{+}H^{-}H^{0}$&
\begin{tabular}{l}
$-\frac{g}{m_{W}}\left[\left(m_{H^{+}}^{2}-m_{A^{0}}^{2}+\frac{1}{2}m_{H^{0}}^{2}\right)\cos\left(\beta-\alpha\right)\right.$
\tabularnewline
$\left.\qquad\quad+\left(m_{A^{0}}^{2}-m_{H^{0}}^{2}\right)\cot2\beta\sin\left(\beta-\alpha\right)\right]$
\end{tabular}\tabularnewline
\hline
$H^{+}H^{-}h^{0}$&
\begin{tabular}{l}
$-\frac{g}{m_{W}}\left[\left(m_{H^{+}}^{2}-m_{A^{0}}^{2}+\frac{1}{2}m_{h^{0}}^{2}\right)\sin\left(\beta-\alpha\right)\right.$
\tabularnewline
$\left.\qquad\quad+\left(m_{h^{0}}^{2}-m_{A^{0}}^{2}\right)\cot2\beta\cos\left(\beta-\alpha\right)\right]$
\end{tabular}\tabularnewline
\hline
$h^{0}h^{0}H^{0}$&
\begin{tabular}{l}
$-\frac{g\cos\left(\beta-\alpha\right)}{2m_{W}\sin2\beta}\left[\left(2m_{h^{0}}^{2}+m_{H^{0}}^{2}\right)\sin2\alpha\right.$
\tabularnewline
$\left.\qquad\qquad\quad-m_{A^{0}}^{2}\left(3\sin2\alpha-\sin2\beta\right)\right]$
\end{tabular}\tabularnewline
\hline
$h^{0}H^{0}H^{0}$&
\begin{tabular}{l}
$-\frac{g\sin\left(\beta-\alpha\right)}{2m_{W}\sin2\beta}\left[\left(2m_{H^{0}}^{2}+m_{h^{0}}^{2}\right)\sin2\alpha\right.$
\tabularnewline
$\left.\qquad\qquad\quad-m_{A^{0}}^{2}\left(3\sin2\alpha+\sin2\beta\right)\right]$
\end{tabular}\tabularnewline
\hline
$A^{0}A^{0}H^{0}$&
\begin{tabular}{l}
$-\frac{g}{2m_{W}}\left[m_{H^{0}}^{2}\cos\left(\beta-\alpha\right)\right.$
\tabularnewline
$\left.\qquad\quad+2\left(m_{H^{0}}^{2}-m_{A^{0}}^{2}\right)\cot2\beta\sin\left(\beta-\alpha\right)\right]$
\end{tabular}\tabularnewline
\hline
$A^{0}A^{0}h^{0}$&
\begin{tabular}{l}
$-\frac{g}{2m_{W}}\left[m_{h^{0}}^{2}\sin\left(\beta-\alpha\right)\right.$
\tabularnewline
$\left.\qquad\quad+2\left(m_{h^{0}}^{2}-m_{A^{0}}^{2}\right)\cot2\beta\cos\left(\beta-\alpha\right)\right]$
\end{tabular}\tabularnewline
\hline
$H^{+}G^{-}H^{0}$&
$-i\frac{g}{2m_{W}}\left(m_{H^{+}}^{2}-m_{H^{0}}^{2}\right)\sin\left(\beta-\alpha\right)$\tabularnewline
\hline
$H^{+}G^{-}h^{0}$&
$i\frac{g}{2m_{W}}\left(m_{H^{+}}^{2}-m_{h^{0}}^{2}\right)\cos\left(\beta-\alpha\right)$\tabularnewline
\hline
$G^{+}G^{-}H^{0}$&
-$i\frac{g}{2m_{W}}m_{H^{0}}^{2}\cos\left(\beta-\alpha\right)$\tabularnewline
\hline
$G^{+}G^{-}h^{0}$&
$-i\frac{g}{2m_{W}}m_{h^{0}}^{2}\sin\left(\beta-\alpha\right)$\tabularnewline
\hline
\end{tabular}
\caption{\label{tab:3h feyn rules}Feynman rules for triple-scalar interactions
\cite{bejar,HHG}.}
\end{table}

\begin{table*}[htb]
\begin{tabular}{|c|c|c|}
\hline
& T2HDM & 2HDM-II \cite{HHG}
\tabularnewline
\hline
$H^{0}\bar{u}_{j}u_{i}$&
$\frac{g}{2m_{W}}\left(-M_{u}\frac{\cos\alpha}{\cos\beta}+\Sigma\left(-\frac{\sin\alpha}{\sin\beta}+\frac{\cos\alpha}{\cos\beta}\right)\right)R+\left(h.c.\right)L$&
$-\frac{gM_{u}}{2m_{W}}\frac{\sin\alpha}{\sin\beta}$\tabularnewline
\hline
$h^{0}\bar{u}u$&
$\frac{g}{2m_{W}}\left(M_{u}\frac{\sin\alpha}{\cos\beta}-\Sigma\left(\frac{\cos\alpha}{\sin\beta}+\frac{\sin\alpha}{\cos\beta}\right)\right)R+\left(h.c.\right)L$&
$-\frac{gM_{u}}{2m_{W}}\frac{\cos\alpha}{\sin\beta}$\tabularnewline
\hline
$A^{0}\bar{u}u$&
$i\frac{g}{2m_{W}}\left(-M_{u}\tan\beta+\Sigma\left(\tan\beta+\cot\beta\right)\right)R+\left(h.c.\right)L$&
$i\frac{gM_{u}}{2m_{W}}\cot\beta\left(R-L\right)$\tabularnewline
\hline
$G^{0}\bar{u}u$&
$i\frac{gM_{u}}{2m_{W}}\left(R-L\right)$&
$i\frac{gM_{u}}{2m_{W}}\left(R-L\right)$\tabularnewline
\hline
$H^{0}\bar{d}d$&
$-\frac{gM_{d}}{2m_{W}}\frac{\cos\alpha}{\cos\beta}$&
$-\frac{gM_{d}}{2m_{W}}\frac{\cos\alpha}{\cos\beta}$\tabularnewline
\hline
$h^{0}\bar{d}d$&
$\frac{gM_{d}}{2m_{W}}\frac{\sin\alpha}{\cos\beta}$&
$\frac{gM_{d}}{2m_{W}}\frac{\sin\alpha}{\cos\beta}$\tabularnewline
\hline
$A^{0}\bar{d}d$&
$i\frac{gM_{d}}{2m_{W}}\tan\beta\left(R-L\right)$&
$i\frac{gM_{d}}{2m_{W}}\tan\beta\left(R-L\right)$\tabularnewline
\hline
$G^{0}\bar{d}d$&
$-i\frac{gM_{d}}{2m_{W}}\left(R-L\right)$&
$-i\frac{gM_{d}}{2m_{W}}\left(R-L\right)$\tabularnewline
\hline
$H^{+}\bar{u}d$&
\begin{tabular}{l}
$\frac{g}{\sqrt{2}m_{W}}\left[\tan\beta V_{CKM}M_{d}R+\right.$\tabularnewline
$\left.+\left(-M_{u}\tan\beta+\Sigma\left(\tan\beta+\cot\beta\right)\right)V_{CKM}L\right]$\tabularnewline
\end{tabular}&
\begin{tabular}{l}
$\frac{g}{\sqrt{2}m_{W}}\left[\tan\beta V_{CKM}M_{d}R+\right.$\tabularnewline
$\left.\qquad+\cot\beta M_{u}V_{CKM}L\right]$\tabularnewline
\end{tabular}\tabularnewline
\hline
$G^{+}\bar{u}d$&
$\frac{g}{\sqrt{2}m_{W}}\left(M_{u}V_{CKM}L-V_{CKM}M_{d}R\right)$&
$\frac{g}{\sqrt{2}m_{W}}\left(M_{u}V_{CKM}L-V_{CKM}M_{d}R\right)$\tabularnewline
\hline
\end{tabular}
\caption{\label{tab:yukawa feyn rules}Feynman rules for Yukawa interactions
in the T2HDM and in the 2HDM-II.}
\end{table*}

\begin{table}[htb]
\begin{tabular}{|c|c|}
\hline
$W^{+}W^{-}H^{0}$&
$igm_{W}\cos\left(\beta-\alpha\right)g^{\mu\nu}$\tabularnewline
\hline
$W^{+}W^{-}h^{0}$&
$igm_{W}\sin\left(\beta-\alpha\right)g^{\mu\nu}$\tabularnewline
\hline
$Z^{0}Z^{0}H^{0}$&
$\frac{igm_{Z}}{\cos\theta_{W}}\cos\left(\beta-\alpha\right)g^{\mu\nu}$\tabularnewline
\hline
$Z^{0}Z^{0}h^{0}$&
$\frac{igm_{Z}}{\cos\theta_{W}}\sin\left(\beta-\alpha\right)g^{\mu\nu}$\tabularnewline
\hline
\end{tabular}
\caption{\label{tab:vvh feyn rules}Feynman rules for vector-vector-scalar
interactions \cite{HHG}.}
\end{table}

The relevant Feynman rules for the 2HDM's that were
used in this work are summarized in Fig. \ref{fig:vertices definition} and in
Tables \ref{tab:yukawa feyn rules}, \ref{tab:vvh feyn rules}, \ref{tab:vhh feyn rules}
and \ref{tab:3h feyn rules}. The notation is given in
Fig. \ref{fig:vertices definition} and the various couplings
in the T2HDM and in the 2HDM of type II are collected in the Tables.

In particular, in Table \ref{tab:yukawa feyn rules} we list the Yukawa couplings,
in Table \ref{tab:vvh feyn rules} we give the vector-vector-scalar couplings, in
Table \ref{tab:vhh feyn rules} we give the vector-scalar-scalar couplings and the triple-scalar
couplings, which are common to any 2HDM \cite{HHG}, are given in Table \ref{tab:3h feyn rules}.
Note that the vertices $Z^{0}G^{0}H^{0}$ and $Z^{0}G^{0}h^{0}$ do not participate
in the calculations since the corresponding Yukawa vertex $\bar{q}qG^{0}$
does not generate FCNC.

\section{\label{app:1L diags}1-loop amplitudes}

Here we give the 1-loop amplitudes corresponding to the 10 diagrams
shown in Fig. \ref{fig: 1-loop diags}. The calculation was done in
the t'Hooft Feynman gauge and the following notation was used:

\textbf{\underbar{definitions:}}

$M_{n}$ -- the amplitude corresponding to diagram $n$.\\
$h$ -- the external neutral scalar.\\
$i$ -- ($=t$) when used as index, the incoming fermion - the top.\\
$j$ -- ($=c$) when used as index, the outgoing fermion - the charm.\\
$\alpha,\beta$ -- when used as indices, internal bosons (vectors or scalars) in the loop.\\
$l,k,q$ -- when used as indices, internal fermions.\\
$L,R$ -- the Left,Right projection operators.\\
$\bar{u}_{j}$ -- ($=\bar{u}(P_{j})$ ) the outgoing spinor of the charm.\\
$u_{i}$ -- ($=u(P_{i})$ ) the incoming spinor of the top.\\
$B_{0},B_{1},C_{0},C_{ij}$ -- the n-point integral functions, defined in App. \ref{app:dijcij definition}.\\
$A_{ab}^{\delta},B_{ab}^{\delta}$ -- the left,right -handed parts of the fermion-fermion-scalar vertex.\\
$a_{ab}^{\delta},b_{ab}^{\delta}$ -- the left,right -handed parts of the fermion-fermion-vector vertex, for both charged and neutral gauge bosons.\\
$g_{abc}^{3h,vhh,vvh}$ -- the vertex of 3-scalars, vector-scalar-scalar, vector-vector-scalar, respectively.

\begin{widetext}
\begin{align}
M_{1} & =\frac{i\bar{u}_{j}}{16\pi^{2}}\frac{-1}{m_{i}^{2}-m_{l}^{2}}\left[
m_{l}m_{k}B_{0}\left(B_{lj}^{h*}A_{lk}^{\alpha}B_{ik}^{\alpha*}L+A_{lj}^{h*}B_{lk}^{\alpha}A_{ik}^{\alpha*}R\right)
-m_{l}m_{i}B_{1}\left(B_{lj}^{h*}A_{lk}^{\alpha}A_{ik}^{\alpha*}L+A_{lj}^{h*}B_{lk}^{\alpha}B_{ik}^{\alpha*}R\right)
\right. \nonumber \\ & \qquad \left.
+m_{i}m_{k}B_{0}\left(B_{lj}^{h*}B_{lk}^{\alpha}A_{ik}^{\alpha*}L+A_{lj}^{h*}A_{lk}^{\alpha}B_{ik}^{\alpha*}R\right)
-m_{i}^{2}B_{1}\left(B_{lj}^{h*}B_{lk}^{\alpha}B_{ik}^{\alpha*}L+A_{lj}^{h*}A_{lk}^{\alpha}A_{ik}^{\alpha*}R\right)\right]u_{i},
\end{align}
where $B=B\left(m_{k}^{2},m_{\alpha}^{2},m_{i}^{2}\right).$
\begin{align}
M_{2} & =\frac{i\bar{u}_{j}}{16\pi^{2}}\frac{-1}{m_{j}^{2}-m_{l}^{2}}\left[
m_{l}m_{k}B_{0}\left(A_{jk}^{\alpha}B_{lk}^{\alpha*}B_{il}^{h*}L+B_{jk}^{\alpha}A_{lk}^{\alpha*}A_{il}^{h*}R\right)
+m_{k}m_{j}B_{0}\left(B_{jk}^{\alpha}A_{lk}^{\alpha*}B_{il}^{h*}L+A_{jk}^{\alpha}B_{lk}^{\alpha*}A_{il}^{h*}R\right)
\right.\nonumber \\ & \qquad \left.
-m_{j}m_{l}B_{1}\left(B_{jk}^{\alpha}B_{lk}^{\alpha*}B_{il}^{h*}L+A_{jk}^{\alpha}A_{lk}^{\alpha*}A_{il}^{h*}R\right)
-m_{j}^{2}B_{1}\left(A_{jk}^{\alpha}A_{lk}^{\alpha*}B_{il}^{h*}L+B_{jk}^{\alpha}B_{lk}^{\alpha*}A_{il}^{h*}R\right)\right]u_{i},
\end{align}
where $B=B\left(m_{k}^{2},m_{\alpha}^{2},m_{j}^{2}\right).$
\begin{align}
M_{3} & =\frac{i\bar{u}_{j}}{16\pi^{2}}\frac{1}{m_{i}^{2}-m_{l}^{2}}\left[
4m_{l}m_{k}B_{0}\left(B_{lj}^{h*}b_{lk}^{\alpha}a_{ik}^{\alpha*}L+A_{lj}^{h*}a_{lk}^{\alpha}b_{ik}^{\alpha*}R\right)
+2m_{l}m_{i}B_{1}\left(B_{lj}^{h*}b_{lk}^{\alpha}b_{ik}^{\alpha*}L+A_{lj}^{h*}a_{lk}^{\alpha}a_{ik}^{\alpha*}R\right)
\right.\nonumber \\ & \qquad \left.
+4m_{i}m_{k}B_{0}\left(B_{lj}^{h*}a_{lk}^{\alpha}b_{ik}^{\alpha*}L+A_{lj}^{h*}b_{lk}^{\alpha}a_{ik}^{\alpha*}R\right)
+2m_{i}^{2}B_{1}\left(B_{lj}^{h*}a_{lk}^{\alpha}a_{ik}^{\alpha*}L+A_{lj}^{h*}b_{lk}^{\alpha}b_{ik}^{\alpha*}R\right)\right]u_i,
\end{align}
where $B=B\left(m_{k}^{2},m_{\alpha}^{2},m_{i}^{2}\right).$
\begin{align}
M_{4} & =\frac{i\bar{u}_{j}}{16\pi^{2}}\frac{1}{m_{j}^{2}-m_{l}^{2}}\left[
4m_{l}m_{k}B_{0}\left(b_{jk}^{\alpha}a_{lk}^{\alpha*}B_{il}^{h*}L+a_{jk}^{\alpha}b_{lk}^{\alpha*}A_{il}^{h*}R\right)
+4m_{k}m_{j}B_{0}\left(a_{jk}^{\alpha}b_{lk}^{\alpha*}B_{il}^{h*}L+b_{jk}^{\alpha}a_{lk}^{\alpha*}A_{il}^{h*}R\right)
\right.\nonumber \\ & \qquad \left.
+2m_{j}m_{l}B_{1}\left(a_{jk}^{\alpha}a_{lk}^{\alpha*}B_{il}^{h*}L+b_{jk}^{\alpha}b_{lk}^{\alpha*}A_{il}^{h*}R\right)
+2m_{j}^{2}B_{1}\left(b_{jk}^{\alpha}b_{lk}^{\alpha*}B_{il}^{h*}L+a_{jk}^{\alpha}a_{lk}^{\alpha*}A_{il}^{h*}R\right)\right]u_i,
\end{align}
where $B=B\left(m_{k}^{2},m_{\alpha}^{2},m_{j}^{2}\right).$

\begin{align}
M_{5} & =\frac{-i\bar{u}_{j}}{16\pi^{2}}\left(A_{jq}^{\alpha}L+B_{jq}^{\alpha}R\right)\left\{
\left[\tilde{C}_{0}+m_{i}^{2}C_{11}+\left(m_{h}^{2}-m_{i}^{2}\right)C_{12}\right]
\left(A_{jq}^{\alpha}A_{kq}^{h*}B_{ik}^{\alpha*}L+B_{jq}^{\alpha}B_{kq}^{h*}A_{ik}^{\alpha*}R\right)
\right.\nonumber \\ & -m_{q}m_{i}C_{11}\left(A_{jq}^{\alpha}B_{kq}^{h*}A_{ik}^{\alpha*}L+B_{jq}^{\alpha}A_{kq}^{h*}B_{ik}^{\alpha*}R\right)
+m_{q}m_{j}C_{12}\left(B_{jq}^{\alpha}A_{kq}^{h*}B_{ik}^{\alpha*}L+A_{jq}^{\alpha}B_{kq}^{h*}A_{ik}^{\alpha*}R\right)
\nonumber \\ & +m_{i}m_{j}\left(C_{12}-C_{11}\right)\left(B_{jq}^{\alpha}B_{kq}^{h*}A_{ik}^{\alpha*}L+A_{jq}^{\alpha}A_{kq}^{h*}B_{ik}^{\alpha*}R\right)
+m_{q}m_{k}C_{0}\left(A_{jq}^{\alpha}B_{kq}^{h*}B_{ik}^{\alpha*}L+B_{jq}^{\alpha}A_{kq}^{h*}A_{ik}^{\alpha*}R\right)
\nonumber \\ & \left.
-m_{i}m_{k}\left(C_{11}+C_{0}\right)\left(A_{jq}^{\alpha}A_{kq}^{h*}A_{ik}^{\alpha*}L+B_{jq}^{\alpha}B_{kq}^{h*}B_{ik}^{\alpha*}R\right)
+m_{j}m_{k}\left(C_{12}+C_{0}\right)\left(B_{jq}^{\alpha}B_{kq}^{h*}B_{ik}^{\alpha*}L+A_{jq}^{\alpha}A_{kq}^{h*}A_{ik}^{\alpha*}R\right)\right\} u_{i},
\end{align}
where $C=C\left(m_{k}^{2},m_{\alpha}^{2},m_{q}^{2},m_{i}^{2},m_{j}^{2},m_{h}^{2}\right).$
\begin{align}
M_{6} & =\frac{i\bar{u}_{j}}{16\pi^{2}}\left\{ \left[4\tilde{C}_{0}+2\left(m_{i}^{2}-m_{j}^{2}+m_{h}^{2}\right)C_{11}+2\left(-m_{i}^{2}+m_{j}^{2}+m_{h}^{2}\right)C_{12}\right]\left(b_{jq}^{\alpha}B_{kq}^{h*}a_{ik}^{\alpha*}L+a_{jq}^{\alpha}A_{kq}^{h*}b_{ik}^{\alpha*}R\right)
\right. \nonumber \\ & +2m_{q}m_{i}C_{11}\left(b_{jq}^{\alpha}A_{kq}^{h*}b_{ik}^{\alpha*}L+a_{jq}^{\alpha}B_{kq}^{h*}a_{ik}^{\alpha*}R\right)
-2m_{q}m_{j}C_{12}\left(a_{jq}^{\alpha}B_{kq}^{h*}a_{ik}^{\alpha*}L+b_{jq}^{\alpha}A_{kq}^{h*}b_{ik}^{\alpha*}R\right)
\nonumber \\ & +4m_{q}m_{k}C_{0}\left(b_{jq}^{\alpha}A_{kq}^{h*}a_{ik}^{\alpha*}L+a_{jq}^{\alpha}B_{kq}^{h*}b_{ik}^{\alpha*}R\right)
+2m_{i}m_{k}\left(C_{11}+C_{0}\right)\left(b_{jq}^{\alpha}B_{kq}^{h*}b_{ik}^{\alpha*}L+a_{jq}^{\alpha}A_{kq}^{h*}a_{ik}^{\alpha*}R\right)
\nonumber \\ & \left.
-2m_{j}m_{k}\left(C_{12}+C_{0}\right)\left(a_{jq}^{\alpha}A_{kq}^{h*}a_{ik}^{\alpha*}L+b_{jq}^{\alpha}B_{kq}^{h*}b_{ik}^{\alpha*}R\right)\right\} u_{i},
\end{align}
where $C=C\left(m_{k}^{2},m_{\alpha}^{2},m_{q}^{2},m_{i}^{2},m_{j}^{2},m_{h}^{2}\right).$
\begin{align}
M_{7} & =\frac{-i\bar{u}_{j}}{16\pi^{2}}g_{\alpha\beta h}^{3h}\left[m_{k}C_{0}\left(A_{jk}^{\beta}B_{ik}^{\alpha*}L+B_{jk}^{\beta}A_{ik}^{\alpha*}R\right)-m_{j}C_{12}\left(B_{jk}^{\beta}B_{ik}^{\alpha*}L+A_{jk}^{\beta}A_{ik}^{\alpha*}R\right)
\right.\nonumber \\ & \left.\qquad\qquad
+m_{i}\left(-C_{11}+C_{12}\right)\left(A_{jk}^{\beta}A_{ik}^{\alpha*}L+B_{jk}^{\beta}B_{ik}^{\alpha*}R\right)\right]u_{i},
\end{align}
where $C=C\left(m_{k}^{2},m_{\alpha}^{2},m_{\beta}^{2},m_{i}^{2},m_{h}^{2},m_{j}^{2}\right).$
\begin{align}
M_{8} & =\frac{-i\bar{u}_{j}}{16\pi^{2}}g_{\alpha\beta h}^{vvh}\left[4m_{k}C_{0}\left(b_{jk}^{\beta}a_{ik}^{\alpha*}L+a_{jk}^{\beta}b_{ik}^{\alpha*}R\right)
+2m_{i}\left(C_{11}-C_{12}\right)\left(b_{jk}^{\beta}b_{ik}^{\alpha*}L+a_{jk}^{\beta}a_{ik}^{\alpha*}R\right)
\right.\nonumber \\ & \left.\qquad\qquad
+2m_{j}C_{12}\left(a_{jk}^{\beta}a_{ik}^{\alpha*}L+b_{jk}^{\beta}b_{ik}^{\alpha*}R\right)\right]u_{i},
\end{align}
where $C=C\left(m_{k}^{2},m_{\alpha}^{2},m_{\beta}^{2},m_{i}^{2},m_{h}^{2},m_{j}^{2}\right).$
\begin{align}
M_{9} & =\frac{i\bar{u}_{j}}{16\pi^{2}}g_{\beta\alpha h}^{vhh}\left[\left(\tilde{C}_{0}+2m_{i}^{2}C_{11}+m_{j}^{2}C_{12}-2m_{h}^{2}C_{12}\right)\left(b_{jk}^{\beta}B_{ik}^{\alpha*}L+a_{jk}^{\beta}A_{ik}^{\alpha*}R\right)
\right.\nonumber \\ & \qquad
-m_{i}m_{j}\left(C_{12}+C_{11}\right)\left(a_{jk}^{\beta}A_{ik}^{\alpha*}L+b_{jk}^{\beta}B_{ik}^{\alpha*}R\right)+m_{j}m_{k}\left(C_{0}-C_{12}\right)\left(a_{jk}^{\beta}B_{ik}^{\alpha*}L+b_{jk}^{\beta}A_{ik}^{\alpha*}R\right)
\nonumber \\ & \left.\qquad
+m_{i}m_{k}\left(C_{12}-C_{11}-2C_{0}\right)\left(b_{jk}^{\beta}A_{ik}^{\alpha*}L+a_{jk}^{\beta}B_{ik}^{\alpha*}R\right)\right]u_{i},
\end{align}
where $C=C\left(m_{k}^{2},m_{\alpha}^{2},m_{\beta}^{2},m_{i}^{2},m_{h}^{2},m_{j}^{2}\right).$
\begin{align}
M_{10} & =\frac{i\bar{u}_{j}}{16\pi^{2}}g_{\alpha\beta h}^{vhh}\left[\left(-\tilde{C}_{0}+m_{i}^{2}\left(C_{12}-C_{11}\right)-2m_{j}^{2}C_{11}-2m_{h}^{2}\left(C_{12}-C_{11}\right)\right)\left(A_{jk}^{\beta}a_{ik}^{\alpha*}L+B_{jk}^{\beta}b_{ik}^{\alpha*}R\right)
\right.\nonumber \\ & \qquad
+m_{i}m_{j}\left(2C_{11}-C_{12}\right)\left(B_{jk}^{\beta}b_{ik}^{\alpha*}L+A_{jk}^{\beta}a_{ik}^{\alpha*}R\right)+m_{j}m_{k}\left(C_{12}+2C_{0}\right)\left(B_{jk}^{\beta}a_{ik}^{\alpha*}L+A_{jk}^{\beta}b_{ik}^{\alpha*}R\right)
\nonumber \\ & \left.\qquad
+m_{i}m_{k}\left(C_{11}-C_{12}-C_{0}\right)\left(A_{jk}^{\beta}b_{ik}^{\alpha*}L+B_{jk}^{\beta}a_{ik}^{\alpha*}R\right)\right]u_{i},
\end{align}
where $C=C\left(m_{k}^{2},m_{\alpha}^{2},m_{\beta}^{2},m_{i}^{2},m_{h}^{2},m_{j}^{2}\right).$
\end{widetext}

\section{\label{app:dijcij definition}1-loop integrals}

The 1-loop scalar, vector and tensor integrals are defined as:
\begin{align}
B_{0};B_{\mu}\left(m_{1}^{2},m_{2}^{2},p^{2}\right) & =\int\frac{d^{4}k}{i\pi^{2}}\frac{1;k_{\mu}}{\left[k^{2}-m_{1}^{2}\right]\left[\left(k+p\right)^{2}-m_{2}^{2}\right]},
\end{align}
\begin{align}
& C_{0};C_{\mu};C_{\mu\nu};\tilde{C}_{0}\left(m_{1}^{2},m_{2}^{2},m_{3}^{2},p_{1}^{2},p_{2}^{2},p_{3}^{2}\right)=
\nonumber \\ & \int\frac{d^{4}k}{i\pi^{2}}\frac{1;k_{\mu};k_{\mu\nu};k^{2}}{\left[k^{2}-m_{1}^{2}\right]\left[\left(k+p_{1}\right)^{2}-m_{2}^{2}\right]\left[\left(k+p_{1}+p_{2}\right)^{2}-m_{3}^{2}\right]},
\end{align}
where $\sum_i p_i =0$ and the reduction to the 1-loop scalar functions is:
\begin{align}
{\rm B}_{\mu} & =p_{\mu}{\rm B}_{1},\nonumber \\
{\rm C}_{\mu} & =p_{1\mu}{\rm C}_{11}+p_{2\mu}{\rm C}_{12},\nonumber \\
{\rm C}_{\mu\nu} & =p_{1\mu}p_{1\nu}{\rm C}_{21}+p_{2\mu}p_{2\nu}{\rm C}_{22}+\{ p_{1}p_{2}\}_{\mu\nu}{\rm C}_{23}+g_{\mu\nu}{\rm C}_{24},
\end{align}
with $\{ ab\}_{\mu\nu}\equiv a_{\mu}b_{\nu}+a_{\nu}b_{\mu}$.

\section{\label{app:higgs width formulae}The Higgs width}

The total Higgs width, $\Gamma^{tot}$, is derived from:
\begin{align}
\Gamma^{tot}= & \Gamma(h\to\bar{q}q)+\Gamma(h\to VV) \nonumber \\
& +\Gamma(h\to H_{i}H_{j})+\Gamma(h\to VH),
\end{align}
when kinematically allowed (i.e.,
the decay products are assumed to be on-shell).
All the above partial widths were calculated at tree-level.
The relevant couplings follow the definition in Fig. \ref{fig:vertices definition}.

The decay width for $h\to\bar{q}q$ is \cite{HHG}:
\begin{align}
\Gamma\left(h\to\bar{q}q\right) & =\frac{N_{c}A_{hqq}^{2}}{8\pi}m_{h}\left(1-\frac{4m_{q}^{2}}{m_{h}^{2}}\right)^{\frac{3}{2}}~,
\end{align}
where $A_{hqq}=-\frac{gm_{q}}{2m_{W}}\frac{\cos\alpha}{\cos\beta}$;
$\frac{gm_{q}}{2m_{W}}\frac{\sin\alpha}{\cos\beta}$ for $h=H^{0};~h^{0}$, respectively, and $N_{c}=3$.
For example, the width for the decay $h^{0}\to\bar{b}b$ in the T2HDM with $\alpha=\beta$ is:
\begin{align}
\Gamma\left(h^{0}\to\bar{b}b\right) & =\frac{3g^{2}m_{b}^{2}}{32\pi m_{W}^{2}}m_{h^{0}}\tan^{2}\beta\left(1-\frac{4m_{b}^{2}}{m_{h^{0}}^{2}}\right)^{\frac{3}{2}}~.
\end{align}

The decay width for $h\to W^{+}W^{-}$ is \cite{HHG}:
\begin{align}
\Gamma\left(h\to W^{+}W^{-}\right) & =\frac{g_{hWW}^{2}m_{h}^{3}}{64\pi m_{W}^{4}}\left(1-x\right)^{\frac{1}{2}}\left(1-x+\frac{3}{4}x^{2}\right)~,
\end{align}
where $g_{hWW}=gm_{W}\cos\left(\beta-\alpha\right)$; $gm_{W}\sin\left(\beta-\alpha\right)$
for $h=H^{0};~h^{0}$, respectively, and
$x=\frac{4m_{W}^{2}}{m_{h}^{2}}$.

The decay width for $h\to Z^{0}Z^{0}$ is \cite{HHG}:
\begin{align}
\Gamma\left(h\to Z^{0}Z^{0}\right) & =\frac{g_{hZZ}^{2}m_{h}^{3}\cos^{4}\theta_{W}}{32\pi m_{W}^{4}}\left(1-x\right)^{\frac{1}{2}}\left(1-x+\frac{3}{4}x^{2}\right),
\end{align}
where $g_{hZZ}=\frac{gm_{Z}}{\cos\theta_{W}}\cos\left(\beta-\alpha\right)$;
$\frac{gm_{Z}}{\cos\theta_{W}}\sin\left(\beta-\alpha\right)$ for
$h=H^{0};~h^{0}$, respectively, and $x=\frac{4m_{Z}^{2}}{m_{h}^{2}}$.
Note that by choosing $\alpha=\beta$ one sets the couplings
$W^{+}W^{-}h^{0}$ and $Z^{0}Z^{0}h^{0}$ to zero in which case
$\Gamma(h^{0}\to W^{+}W^{-}),~\Gamma(h^{0}\to Z^{0}Z^{0}) =0$ while
$\Gamma(H^{0}\to W^{+}W^{-})$ and $\Gamma(h^{0}\to Z^{0}Z^{0})$ are maximal.

The decay width for $h\to H_{i}H_{j}$ (where $H_{i},~H_{j} \neq h$) is:
\begin{align}
\Gamma\left(h\to H_{i}H_{j}\right) & =\frac{g_{hH_{i}H_{j}}^{2}}{16\pi m_{h}}\lambda^{\frac{1}{2}}\left(1,\frac{m_{H_{i}}^{2}}{m_{h}^{2}},\frac{m_{H_{j}}^{2}}{m_{h}^{2}}\right)~,
\end{align}
where $g_{hH_{i}H_{j}}$ are the triple scalar couplings, and we recall that: $\lambda\left(x,y,z\right)=x^{2}+y^{2}+z^{2}-2xy-2xz-2yz$.

The decay width for $h\to V H$ (where $V H=W^{+}H^{-}$ or
$VH= Z^{0}H_i^0$ and $H_i^0 \neq h$) is:
\begin{align}
\Gamma\left(h\to VH\right) & =\frac{g_{VHh}^{2}m_{V}^{2}}{16\pi m_{h}}\lambda^{\frac{1}{2}}\left(1,\frac{m_{V}^{2}}{m_{h}^{2}},\frac{m_{H}^{2}}{m_{h}^{2}}\right)
\lambda\left(1,\frac{m_{h}^{2}}{m_{V}^{2}},\frac{m_{H}^{2}}{m_{V}^{2}}\right)~,
\end{align}
where $g_{hVH}$ are the vector-scalar-scalar couplings.

\begin{figure}
\begin{centering}
\includegraphics[width=8cm,keepaspectratio]{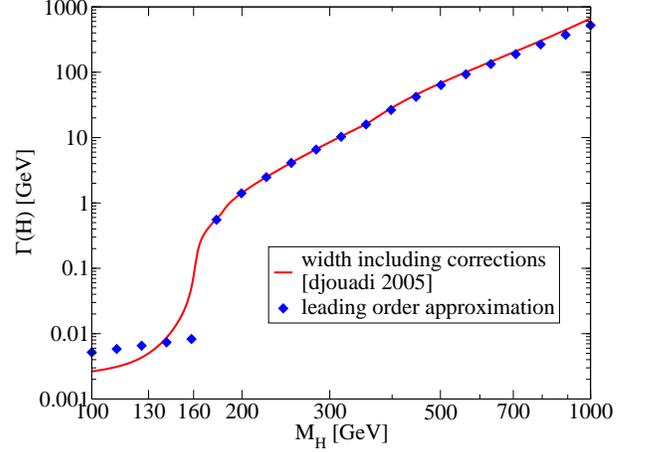}
\par\end{centering}
\caption{\label{fig:higgs width SM}The total width of the SM Higgs: leading
order approximation compared to the corrected (1-loop) width of \cite{djouadi I}.}
\end{figure}

In order to demonstrate the role of radiative corrections to the
leading order tree-level total width, we plot in Fig. \ref{fig:higgs width SM}
the total SM Higgs width at the tree-level (i.e., as calculated in this work)
compared to the width calculated including higher-order corrections \cite{djouadi I}.
As can be seen, the discrepancy between the lowest order and the higher
order calculations is significant only below the WW threshold (at about 160 GeV).
In this mass range the $b\bar{b}$ decay channel dominates for which radiative corrections
can have an appreciable impact. This mass range is, however,
below the $h\to\bar{t}c$ threshold and therefore irrelevant to the present
work, and so the use of the lowest order widths is justified.

\begin{thebibliography}{10}

\bibitem{Das}A. Das, C. Kao, {}``A two Higgs doublet model for the
top quark'', Phys. Lett. B\textbf{372}, 106 (1996), arXiv:hep-ph/9511329.

\bibitem{bejar}S. Bejar, {}``Flavor changing neutral decay effects
in models with two Higgs boson doublets: Applications to LHC Physics'',
PhD thesis (2006), arXiv:hep-ph/0606138; S. Bejar, J. Guasch, J. Sola,
{}``Loop Induced Flavor Changing Neutral Decays of the Top Quark
in a General Two-Higgs-Doublet Model'', Nucl. Phys. B\textbf{600},
21 (2001), arXiv:hep-ph/0011091.

\bibitem{t-ch LHC aguilar-saavedra}J.A. Aguilar-Saavedra, ''Top
flavor-changing neutral interactions: Theoretical expectations and
experimental detection'', Acta Phys. Polon. B\textbf{35}, 2695 (2004),
arXiv:hep-ph/0409342; J.A. Aguilar-Saavedra, G.C. Branco, {}``Probing
top flavor changing neutral scalar couplings at the CERN LHC'', Phys.
Lett. B\textbf{495}, 347 (2000), arXiv:hep-ph/0004190.

\bibitem{t-ch SM}
B. Grzadkowski, J.F. Gunion, P. Krawczyk, {}``Neutral Current Flavor Changing Decays for the Z Boson and the Top Quark in Two Higgs Doublet Models'', Phys. Lett. B\textbf{268}, 106-11 (1991), UCD-90-34, Dec 1990; 
J.L. Diaz-Cruz, R. Martinez,M.A. Perez and A. Rosado, {}``Flavor Changing Radiative Decay of the Top", Phys. Rev. D\textbf{41}, 891 (1900); 
G. Eilam, J.L. Hewett, A. Soni, {}``Rare decays of the top quark in the standard and two Higgs doublet models'', Phys. Rev. D\textbf{44}, 1473 (1991), see also: Erratum, Phys. Rev. D\textbf{59}:039901 (1999); 
B. Mele, S. Petrarca, A. Soddu, ''A New evaluation of the $t\to cH$ decay width in the standard model'', Phys. Lett. B\textbf{435}, 401 (1998), arXiv:hep-ph/9805498.

\bibitem{h-tc SM arhrib}A. Arhrib, {}``Higgs bosons decay into bottom-strange
in two Higgs Doublets Models'', Phys. Lett. B\textbf{612}, 263 (2005),
arXiv:hep-ph/0409218.

\bibitem{HHG}J. F. Gunion, H. E. Haber, G. Kane, S. Dawson, {}``The
Higgs Hunter's Guide'', Addison-Wesley (1990); see also: Errata,
SCIPP-92-58 (1992), arXiv:hep-ph/9302272.

\bibitem{georgi soft CP and Z2}H. Georgi, {}``A model of soft CP
violation'', Hadronic J. \textbf{1}, 155 (1978).

\bibitem{soni 15 Z-bs}D. Atwood, S. Bar-Shalom, G. Eilam, A. Soni,
''Flavor changing Z-decays from scalar interactions at a Giga-Z Linear
Collider'', Phys. Rev. D\textbf{66}:093005 (2002), arXiv:hep-ph/0203200.

\bibitem{soni 21 most calcs}G.-H. Wu, A. Soni, {}``Novel CP-violating
effects in B decays from a charged Higgs boson in a two-Higgs-doublet
model for the top quark''. Phys. Rev. D\textbf{62}:056005 (2000),
arXiv:hep-ph/9911419.

\bibitem{atwood reina soni bound neutral}D. Atwood, L. Reina, A.
Soni, ``Phenomenology of two Higgs doublet models with flavor changing
neutral currents'', Phys. Rev. D\textbf{55}, 3156 (1997), arXiv:hep-ph/9609279.

\bibitem{soni 13 3jet}D. Atwood, S. Bar-Shalom, G. Eilam, A. Soni,
``Three heavy jet events at hadron colliders as a sensitive probe
of the Higgs sector'', Phys. Rev. D\textbf{69}:033006 (2004), arXiv:hep-ph/0309016.

\bibitem{thesis}I. Baum, ``Top quark rare decays in a two Higgs doublet model for the top", MSc Thesis (2007), arXiv:hep-ph/0711.1311.

\bibitem{soni 34 best}E. Lunghi, A. Soni, {}``Footprints of the
Beyond in flavor physics: Possible role of the Top Two Higgs Doublet
Model'', J. of High Energy Physics \textbf{0709}:053 (2007), 
arXiv:hep-ph/0707.0212.

\bibitem{arhrib}A. Arhrib, {}``Top and Higgs Flavor Changing Neutral
Couplings in two Higgs Doublets Model'', Phys. Rev. D\textbf{72}:075016
(2005), arXiv:hep-ph/0510107.

\bibitem{PDG}W.-M. Yao et al. (Particle Data Group), J. Phys. G\textbf{33},
1 (2006) and 2007 partial update for the 2008 edition (URL: http://pdg.lbl.gov).

\bibitem{peskin}M. E. Peskin, D. V. Schroeder, \char`\"{}An introduction
to quantum field theory\char`\"{}, Perseus Books (1995).

\bibitem{ff vanold}G. J. van Oldenborgh, {}``FF: A Package to evaluate
one loop Feynman diagrams'', NIKHEF-H-90-15 (1990); Comput. Phys.
Commun. \textbf{66}, 1 (1991); download: http://www.xs4all.nl/\textasciitilde{}gjvo/FF.html.

\bibitem{djouadi II}A. Djouadi, ``The Anatomy of Electro-Weak Symmetry
Breaking, II: The Higgs bosons in the Minimal Supersymmetric
Model'', LPT-ORSAY-05-18 (2005), arXiv: hep-ph/0503173.

\bibitem{running masses 2000}C.R. Das, M.K. Parida, ``New formulae
and predictions for running fermion masses at higher scales in SM,
2 HDM, and MSSM'', Eur. Phys. J. C\textbf{20}, 121 (2001), arXiv:hep-ph/0010004.

\bibitem{djouadi I}A. Djouadi, ``The Anatomy of Electro-Weak Symmetry
Breaking, I: The Higgs boson in the Standard Model'', LPT-ORSAY-05-18
(2005), arXiv:hep-ph/0503172.

\end{thebibliography}
\end{document}